\documentclass[showpacs,aps,nofootinbib,superscriptaddress, twocolumn]{revtex4}
\usepackage{natbib}
\usepackage[utf8]{inputenc}
\usepackage{graphics}
\usepackage{amssymb}
\usepackage{bm}
\usepackage{braket}
\usepackage{booktabs}
\usepackage{subfloat}
\usepackage{float}

\usepackage{amsmath}    
\usepackage{graphicx}   
\usepackage{verbatim}   
\usepackage{color}      
\usepackage{subfigure}  
\usepackage{hyperref}   
\usepackage{algorithm,algorithmic}

\hypersetup{
    colorlinks=true,
    linkcolor=red,
    citecolor=blue,
}

\begin{document}
\title{Diversity metric for evaluation of quantum annealing}


\author{Alex Zucca}
\affiliation{D-Wave Systems Inc.}
\affiliation{Simon Fraser University}
\author{Hossein Sadeghi}
\affiliation{D-Wave Systems Inc.}
\author{Masoud Mohseni}
\affiliation{Google Inc.}
\author{Mohammad H. Amin}
\affiliation{D-Wave Systems Inc.}
\affiliation{Simon Fraser University}
\date{\today}

\begin{abstract}
Solving discrete NP-hard problems is an important part of scientific discoveries and operations research as well as many commercial applications. A commonly used metric to compare meta-heuristic solvers is the time required to obtain an optimal solution, known as time to solution. However, for some applications it is desirable to have a set of high-quality and diverse solutions, instead of a single optimal one. For these applications, time to solution may not be informative of the performance of a solver, and another metric would be necessary. In particular, it is not known how well quantum solvers sample the configuration space in comparison to their classical counterparts.
Here, we apply a recently introduced collective distance measure in solution space to quantify diversity by Mohseni, et. al. \cite{2021arXiv211010560M} and, based on that, we employ \textit{time-to-diversity} as a metric for evaluation of meta-heuristics solvers.
We use this measure to compare the performance of the D-Wave quantum annealing processor with a few classical heuristic solvers on a set of synthetic problems and show that D-Wave quantum annealing processor is indeed a competitive heuristic, and on many instances outperforms state-of-the-art classical solvers, while it remains on par on other instances. This suggests that a portfolio solver that combines quantum and classical solutions may win over all solvers.

\end{abstract}
\maketitle

\vspace{-0.5cm}
\section{Introduction}
\label{sec:intro}
The development of quantum annealing (QA) processors  \citep{M.W.Johnson2011Qawm}, has sparked a renewed interest on  combinatorial optimization for industrial problems. The potential advantage of QA processors lies in the fact that they have access to quantum resources, such as superposition and entanglement among variables, which could lead to the possibility of non-trivial quantum tunneling or collective moves in configuration space that are absent in local stochastic classical heuristics. The quantum annealer is usually tested against some classical algorithms, such as simulated annealing (SA), parallel tempering (PT), and parallel tempering with iso-energetic cluster moves (PT+ICM) \cite{2015PhRvL.115g7201Z, 2016arXiv160509399Z}, with the general trend that QA competes fairly well with the state of the art heuristics.
Due to the limitations of the current QA processors in terms of number of variables and their inter-connectivity, the search for quantum speed-up is usually performed on a set of synthetic benchmark problems  \cite{2015PhRvA..92d2325H, 2015arXiv150202098K, 2017arXiv170104579K, 2018PhRvX...8c1016A, 2018QS&T....3dLT01M}, where the ``hardness'' of the problems is, in some cases, tuneable via one or more parameters. 
However, despite many attempts to detect quantum speedup in the current generation of quantum annealers over classical algorithms \cite{DicksonNG2013Taqa, SergioBoixo2014Efqa, KatzgraberHelmutG.2015SQST, HeimBettina2015Qvca, 2014PhRvX...4b1008K, 2015PhRvX...5c1026K, 2015Sci...348..215H, 2015PhRvA..92d2325H, 2015EPJST.224..111A, 2015NatSR...515324M, 2015arXiv150202098K, 2016PhRvA..94a2320M, 2016PhRvX...6c1015D, 2017arXiv170104579K, 2018PhRvX...8c1016A}, the existence of an improved scaling in optimization still remains elusive, although such a scaling advantage has recently been reported in quantum simulation of magnetic materials \cite{2021NatCo..12.1113K}. 


Most previous studies reported the performances in terms of time-to-solution (TTS), typically defined as the total time (which could be the sum of several repetitions) required to reach an optimal solution with 99\% certainty. It is clear that in order to calculate TTS, knowledge of the optimal solution is necessary, which can become intractable for large problems. Moreover, TTS is defined based on obtaining a single solution and therefore does not give any information on how widely the returned solutions are distributed in the configuration space. Other metrics such as the Time-To-Target (TTT) introduced in \cite{2015arXiv150805087K} are designed to provide a benchmarking quantity that does not become intractable as the number of variables increases. This is achieved by considering sub-optimal solutions provided by a reference solver ({\it e.g.} the quantum annealer) and estimating how long other solvers take to achieve the same level of sub-optimality. Other studies instead, analyzed the sampling abilities of QA. In particular it was noted in~\cite{2009JPhCS.143a2003M} that transverse field driver Hamiltonians like the one implemented in current quantum annealers result in an unfair sampling of degenerate ground states. Indeed this was confirmed in a systematic study of the D-Wave 2X annealer ~\cite{2017PhRvL.118g0502M} and it was suggested the more complex driver Hamiltonian may help reduce the bias, although with the conditions of being dense enough \cite{2018arXiv180606081K}. Nevertheless, QA can be complementary to classical algorithms to obtain ``classically rare'' solutions \cite{2017NatSR...7.1044Z}. Recently, fair sampling performances of state-of-the-art classical algorithms such as PT and PT+ICM were put under scrutiny  \cite{2019PhRvE..99f3314Z}. Although it emerged that PT+ICM performs better than PT, it was also noted the performances degrade as the Hamming distance between ground states increase. 
A different point of view was taken in \cite{doi:10.7566/JPSJ.88.061007} where the authors defined the Time-To-All-Valleys (TTAV) metric to measure the ability of a solver to obtain at least one sample in an island of ground states, where the latter is defined as a configuration of ground states that are connected in the logical Hamming space. Since moving within the same valley can be done with minimal post-processing, being able to reach quickly all the valleys is of critical importance. 
 
 In this work we adopt a similar point of view, but we do not restrict to diverse ground state configurations. Rather, we focus on the ability of solvers to reach diverse sub-optimal solutions.
In fact, due to noise, calibration errors, and finite temperature, QA processors are expected to produce a pool of  diverse,  high-quality but sub-optimal solutions.
Diversity of high-performing population is of crucial importance in genetic programming, evolutionary algorithms, quality-diversity optimization \cite{10.3389/frobt.2016.00040}, contingency planning, robust optimization, and machine learning applications \cite{10.1145/1830483.1830569, 10.1145/2001576.2001665,  2015arXiv151008568G, 10.1145/3071178.3071342, 2018arXiv180205448N}. 
Recently, a new diversity measure was introduced based on the number of high-quality solutions, within a target approximation ratio, that have large mutual Hamming distance larger than a certain threshold \cite{2021arXiv211010560M}. Here, we adopt such diversity measure to quantify the performance of the D-Wave 2000Q annealer in terms of time-to-diversity (TTD) - the time the solver takes to return a sample set with target diversity. 

The paper is structured as follows. In Sect.~\ref{sec:optimization} we review how the D-Wave quantum processing unit (QPU) solves an Ising minimization problem. In Sect.~\ref{sec:diversity} we review the Hamming-distance-based \emph{diversity measure} that can be used to characterize the ability of a solver to return diverse samples \cite{Mohseni:2021}. Similarly to the definition of TTS, we employ the concept of TTD and we introduce two ways of obtaining its upper and lower bounds. We present and discuss the results in Sect.~\ref{sec:results_and_discussion}. We then conclude in Sect.~\ref{sec:conclusions}.

\section{Overview of the D-Wave hardware}
\label{sec:optimization}
The D-Wave quantum annealer is designed to solve Ising-like binary optimization problems, and implements, in the close system approximation, the following Hamiltonian,
\begin{align}
\label{eqn:DWAVEHamiltoniainCS}
H & = - A(s)  \sum_i \sigma_i^x+ B(s) \mathcal{H}_P, \\
\mathcal{H}_P & = \sum_i h_i \sigma_i^z + \sum_{i,j} J_{ij}\sigma_i^z \sigma_j^z,
\end{align}
where $h_i$ and $J_{ij}$ are local and pairwise parameters, respectively, and $\sigma_i^x$ ($\sigma_i^z$) represents the Pauli matrices in $x$ ($z$) direction. The set of indices $(i,j)$ forms the adjacency matrix of the Chimera graph, the  architecture of the D-Wave 2000Q quantum annealer \cite{2014ITAS...2418294B}.
The two  functions $A(s)$ and $B(s)$ -- known as the annealing schedule -- have a pre-defined time dependent form expressed in terms of the normalized time $s = t/t_a$,  with $t_a $ being the total anneal time.  
A problem of interest is usually mapped (in polynomial time) \cite{2014FrP.....2....5L} to the problem of minimizing the energy of an Ising Hamiltonian of the form above.\footnote{While mapping a NP-hard problem to an Ising Hamiltonian is polynomial, finding the optimal embedding onto the Chimera graph is NP-hard itself. More time-efficient heuristic ways to find an embedding are used instead  \cite{2016QuIP...15..495B, 2014arXiv1406.2741C}.}
In forward annealing, the system is initially prepared in the ground state of the Hamiltonian when $A(0) \gg B(0)$.  The energy scale of the driver Hamiltonian -- $A(s)$ -- is then slowly reduced, while the energy scale of the problem Hamiltonian -- $B(s)$ -- is increased to give, at end of the anneal,  $B(1) \gg A(1) \approx 0$. At $s=1$,  the Hamiltonian \eqref{eqn:DWAVEHamiltoniainCS} reduces to the classical energy of the Ising model, at which point we replace the Pauli matrices with the values of spin represented by $s_i \in \{\pm 1\}$, 
\begin{equation}
\label{eqn:IsingHamiltonian}
   \mathcal{H}_P =  \sum_i h_i s_i + \sum_{i,j} J_{ij}s_i s_j.
\end{equation}
If the process is done slowly enough, due to the adiabatic theorem \cite{andp.19163561905, born_fock}, the system will remain in the ground state of the instantaneous Hamiltonian, and will  return at the end of the anneal the spins configuration that minimizes the Ising Hamiltonian \eqref{eqn:IsingHamiltonian}.

In practice, however, the evolution is usually non-adiabatic and is done in the presence of an environment with finite temperature and Hamiltonian parameter specification error. As such, instead of obtaining the ground state configuration at every anneal, one rather obtains a distribution of configurations with a probability $p$ to obtain a ground state solution. 
Here we turn our attention on a benchmarking quantity that does not involve the achievement of the optimal solution. Rather, we assess the ability to explore sub-optimal solutions in a different manner. While estimating the solutions to a sub-optimal point is proven to be polynomial in time \cite{2013arXiv1306.6943S}, the pre-factor of such algorithms grows exponentially with the inverse of normalized residual energy, turning these methods impractical for solving problems of our interest.
\section{Diversity measure based on mutual Hamming distance}
\label{sec:diversity}
In this Section, we define the \emph{diversity measure} based on the mutual Hamming distance between different samples \cite{Mohseni:2021}, and adopt the related idea of Time-To-Diversity (TTD). Because the computation of diversity becomes quickly intractable with the system size, we identify some more efficient strategies to assess the diversity of the samples returned by solvers.

Let us consider the set of $2^N$ configurations of a binary optimization problem with Hamiltonian \eqref{eqn:IsingHamiltonian} with $N$ being the total number of spins, and let us denote the ground state energy as $E_0$. To restrict our attention to sub-optimal configurations, for a given solution ${\bf s}^a = \{ s^a_i \}$ with energy $E({\bf s}^a)$ we define {\em approximation ratio} as
\begin{equation}
\label{eqn:approximation_ratio}
r({\bf s}^a) = \frac{E({\bf s}^a) - E_0}{E_{\rm max} - E_0},
\end{equation}
where $E_{\rm max}$ is the maximum energy of the problem Hamiltonian. We choose a threshold value $ 0 \le \alpha \ll 1 $ for the approximation ratio and a corresponding energy threshold 
\begin{equation}
\label{eqn:energy_threshold}
E_{\alpha} = E_0 + \alpha (E_{\rm max} - E_0).
\end{equation}
Let us define $\mathcal{S} = \{{\bf s}^1, \dots, {\bf s}^M \}$ as a set of $M$ spin configurations ${\bf s}^a$ having $E_0 \leq E({\bf s}^a) \leq E_{\alpha}$.
For all states ${\bf s}^a\in \mathcal{S}$, we can build the Hamming distance matrix
\begin{equation}
d_{ab} = \frac{1}{2}\sum_{i=1}^N |s^a_i - s^b_i|,
\end{equation}
where $a$ and $b$ are the indices of solutions and $i$ is the index of individual binary units (qubits).
We then construct a graph $\mathcal{G}_{\mathcal{S}  R}=(\mathcal{V}=\mathcal{S}, \mathcal{E})$ where each solution ${\bf s}^a \in \mathcal{S}$ is a node, and the edges are defined only for the pairs of solutions ${\bf s}^a, {\bf s}^b$ that have Hamming distance smaller than a normalized radius $R$ through $d_{ab} \leq R N$. An independent set $V_{\rm IS}$ is a set of vertices, no pair of which are connected by an edge. The \emph{diversity} $D_{\mathcal{S} R}$ is defined as the \emph{independence number} of the graph $\mathcal{G}_{\mathcal{S} R}$, {\it i.e.} the size of the maximum independent set (MIS)
\begin{equation}
D_{\mathcal{S} R} = \max_{V_{\rm IS} \subseteq \mathcal{V}}|V_{\rm IS}|.
\end{equation}
Note that $E_{\alpha}$ is chosen based on the approximation ratio that requires knowledge of the ground state energy. However, it is possible to choose the energy threshold directly (e.g., based on the application) without knowing the ground state and the following discussions remain unaffected. 
Exact calculation of $D_{\mathcal{S} R}$ becomes computationally intractable when the number of solutions in $\mathcal{S}$ is large. Nevertheless, we can compute approximated diversity as we discuss the following.

\subsection{Approximated Diversity Measures}
Here, we show that upper and lower bounds on the diversity measure can be computed with polynomial algorithms. For the lower bound, $D_{\mathcal{S}R}^{\rm low}$ we use a recursive algorithm that we call \emph{Left Neighbors Aggregation} (LNA) algorithm. The algorithm sequentially scans every vertex and add it to the current independent set if the vertex is not connected to the previous members of the set. Naturally, the order with which the samples are scanned is important, and to tighten the bound we analyze many random orders for the set $\mathcal{S}$. The algorithm is described in details in App.~\ref{sec:lna_algorithm}. 
The LNA algorithm complexity is $\mathcal{O}(M \cdot D_{\mathcal{S}R}^{\rm low})$ . In practical cases $D_{\mathcal{S}R}^{\rm low}\ll M$, so that the LNA scales almost linearly with the number of samples $M$.

The upper bound $D_{\mathcal{S} R}^{\rm up}$ is computed by approximating the \emph{chromatic number} $\chi$ of the complement graph $\bar{\mathcal{G}}_{\mathcal{S} R}$ \footnote{The chromatic number $\chi(\mathcal{G})$ of a graph $\mathcal{G}$ is the minimum $k$ for which $\mathcal{G}$ admits a \emph{legal} coloring of it vertices. A legal coloring of a graph is an assignment of colors to its vertices so that no pair of adjacent vertices  has the same color. The chromatic number is also an upper bound on the clique number $\omega(\mathcal{G})$ of the same graph, which is equivalent to the independence number $\alpha(\bar{G})$ of the complement graph $\bar{\mathcal{G}}$, $\alpha(\bar{\mathcal{G}}) = \omega(\mathcal{G}) \le \chi(\mathcal{G})$.} with a greedy coloring algorithm \cite{Kosowski2004ClassicalCO, Matula:1983:SOC:2402.322385}. Computing the approximated chromatic number with the greedy algorithm is $\mathcal{O}(M^2)$ in the number of samples $M$, as one needs to compute the distance matrix $d_{ab}$ in advance. The greedy coloring algorithm is described in App.~\ref{sec:greedy_coloring}.

As we show in App.~\ref{sec:algorithm_performances}, the lower bound obtained by the LNA algorithm is tighter than the upper bound and scales similarly as the independence number. Thus, in the following we will exclusively refer to diversity as its lower bound computed with the LNA algorithm.

\subsection{Time-to-Target-Diversity}
\label{sec:practical_target_diversity}
Inspired by \cite{2015arXiv150805087K} we introduce a strategy to compare the diversity performance of different solvers.
We consider one of the solvers as a reference solver $\kappa_r$ - in our case the D-Wave QPU - and define a target diversity in the following way. We obtain a fixed number of samples from the reference solver, then remove all samples with energy above the threshold $E_{\alpha}$. This new set of samples, denoted as $\mathcal{S}_{\alpha}^{r}$, is then used to calculate the diversity. For a fixed approximation ratio $\alpha$ and hamming distance radius $R$, the target diversity is defined as $\hat{D} \equiv D_{\mathcal{S}_{\alpha}^r R}$.

Consider a set of solutions $\mathcal{S}_{\kappa}(t)$ with energies below $E_{\alpha}$,  returned by the solver $\kappa$ within finite time $t = t_a n_{\rm runs}$, where $n_{\rm runs}$ is the number of repetitions and $t_a$ is the time it takes to run the algorithm once.
For simplicity of notation, we denote the diversity of this set as $D_{\kappa}(t) \equiv D_{\mathcal{S}_{\kappa}(t) R}$.
Let $p_\kappa$ be the probability of the solver $\kappa$ reaching the target diversity, $D_{\kappa}(t) \ge \hat{D}$, after time $t$. Similarly to TTS, we define Time-to-Target-Diversity (TTD) as the time it takes for the solver to reach the target diversity $\hat{D}$ with $99 \%$ confidence,
\begin{equation}
\label{eqn:ttd_definition}
{\rm TTD}_\kappa = n_{\rm runs} t_a \frac{\log(0.01)}{\log(1 - p_\kappa)}.
\end{equation}

\section{Results and discussion}
\label{sec:results_and_discussion}
In this Section we compare the DW2000Q quantum annealer against two commonly used classical solvers, namely Parallel Tempering (PT) and Parallel Tempering with iso-energetic cluster moves (PT+ICM). 
In our benchmark we  consider the reference diversity $\hat{D}$ using 50k samples obtained by the DW2000Q solver. 

We begin our analysis using an approximation ratio $\alpha=0.01$ and a radius factor $R=0.25$. Smaller approximation ratios require significant computation times. We also expect the performance of the QPU to degrade due to noise and calibration error. Nevertheless, we studied $\alpha=0.005$ with smaller number of problems and observed similar behavior as shown in Sect.~\ref{sec:small_alpha}.


\subsection{Input Classes}
We analyzed ${\rm TTD}$ for three different input classes. Each of them is randomly generated on the underlying Chimera graph of the available D-Wave QPU. The inputs were generated using only the first $L\times L$ unit cells of the Chimera graph with $L \in \{ 8, 12, 16 \}$ with 512, 1152 and 2048 qubits respectively. All the classes have linear terms $h_i$ set to zero, so that
\begin{equation}
    \mathcal{H}_P = \sum_{(i, j) \in C_L} J_{ij} s_i s_j,
\end{equation}
with $C_L$ denoting available connectivity graph of an D-Wave QPU. The input classes are:

\begin{itemize}
    \item {\bf RAN1 - Bimodal spin glasses.} These inputs are generated by randomly assigning each coupling a value in $\{\pm 1 \}$ with equal probability.
    \item {\bf AC3 - Anti-cluster inputs.} These inputs are RAN1 problems with couplings between unit cells  multiplied by 3.
    \item {\bf DCL - Deceptive Cluster Loops}. The DCL problems \cite{2018QS&T....3dLT01M} are a variant of the frustrated cluster loops (FCL) \cite{2017arXiv170104579K}. Here we set the parameters of the DCL problems to $(\alpha_{\rm DCL}, R_{\rm DCL}, \lambda) = (0.25, 1, 7)$.
\end{itemize} 

For each  size  $L$ we generated 50 instances of each RAN1, AC3 and DCL type.
\begin{figure*}[tbh]
    \centering
    \subfigure{
    \includegraphics[width=.32\textwidth]{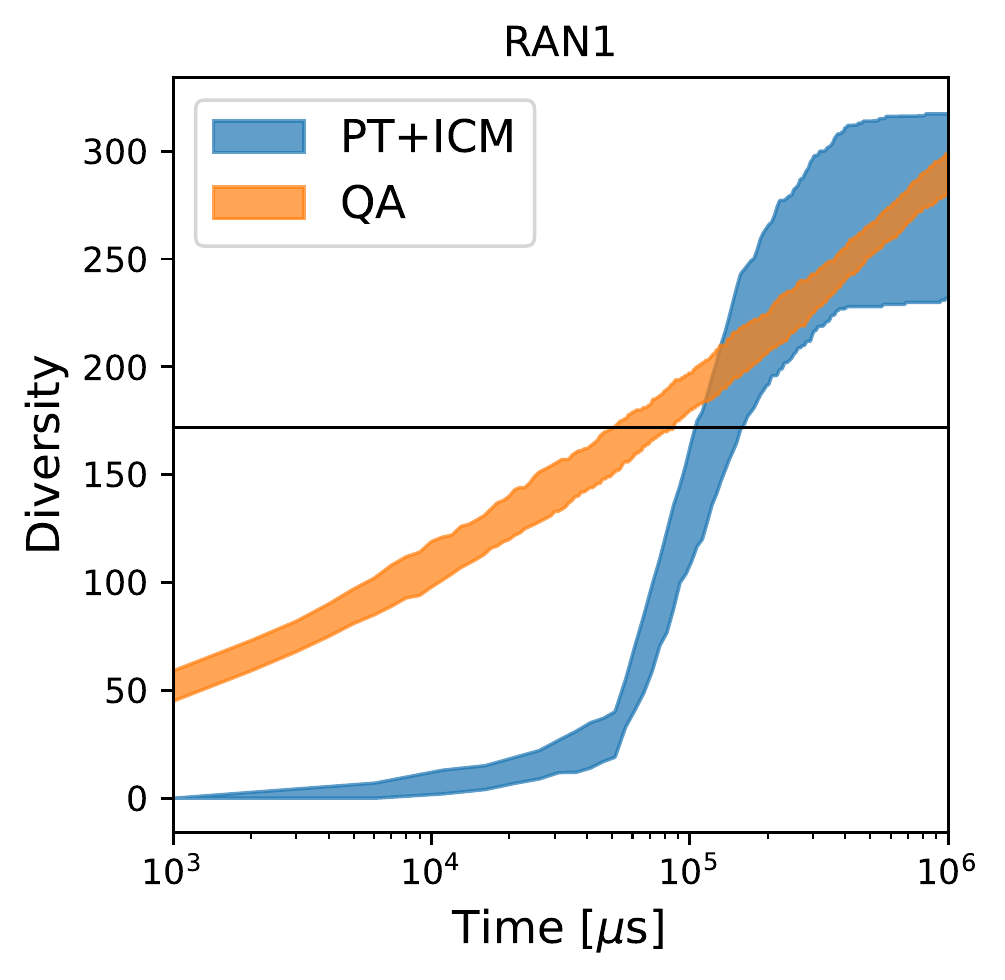}}
    \subfigure{
    \includegraphics[width=.32\textwidth]{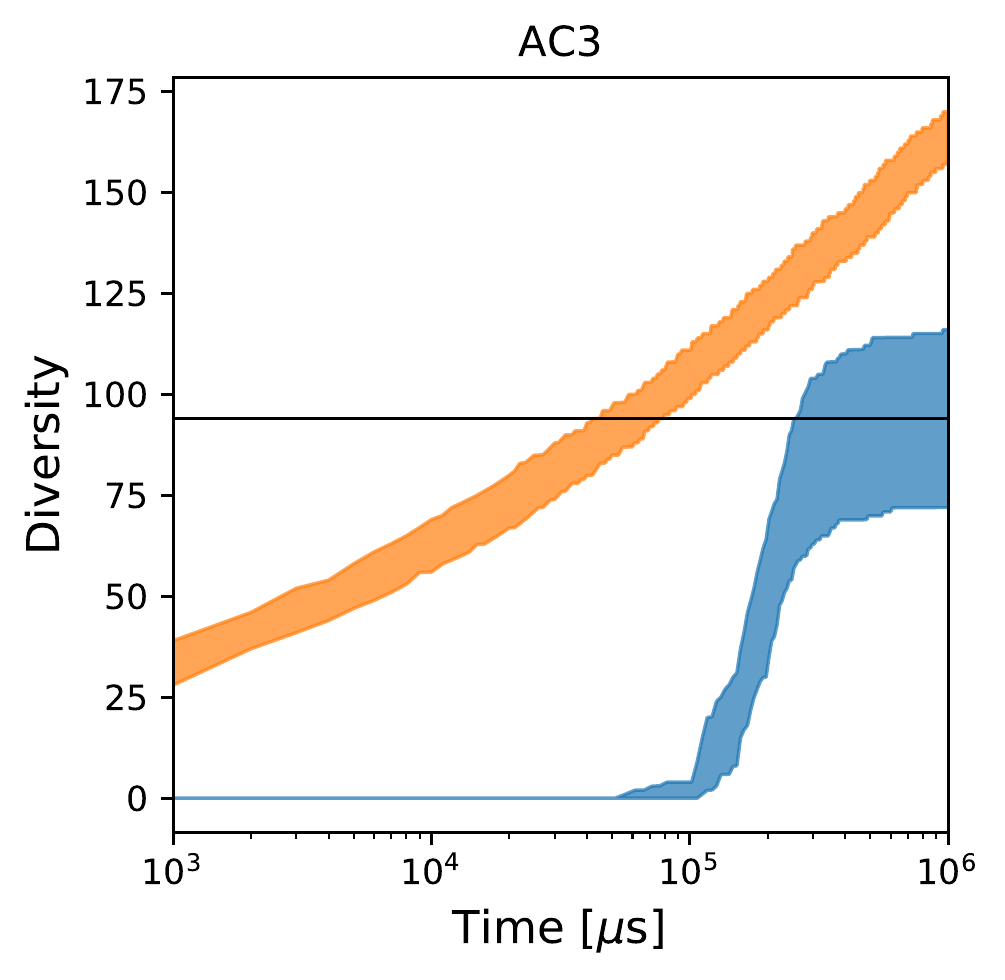}}
    \subfigure{
    \includegraphics[width=.32\textwidth]{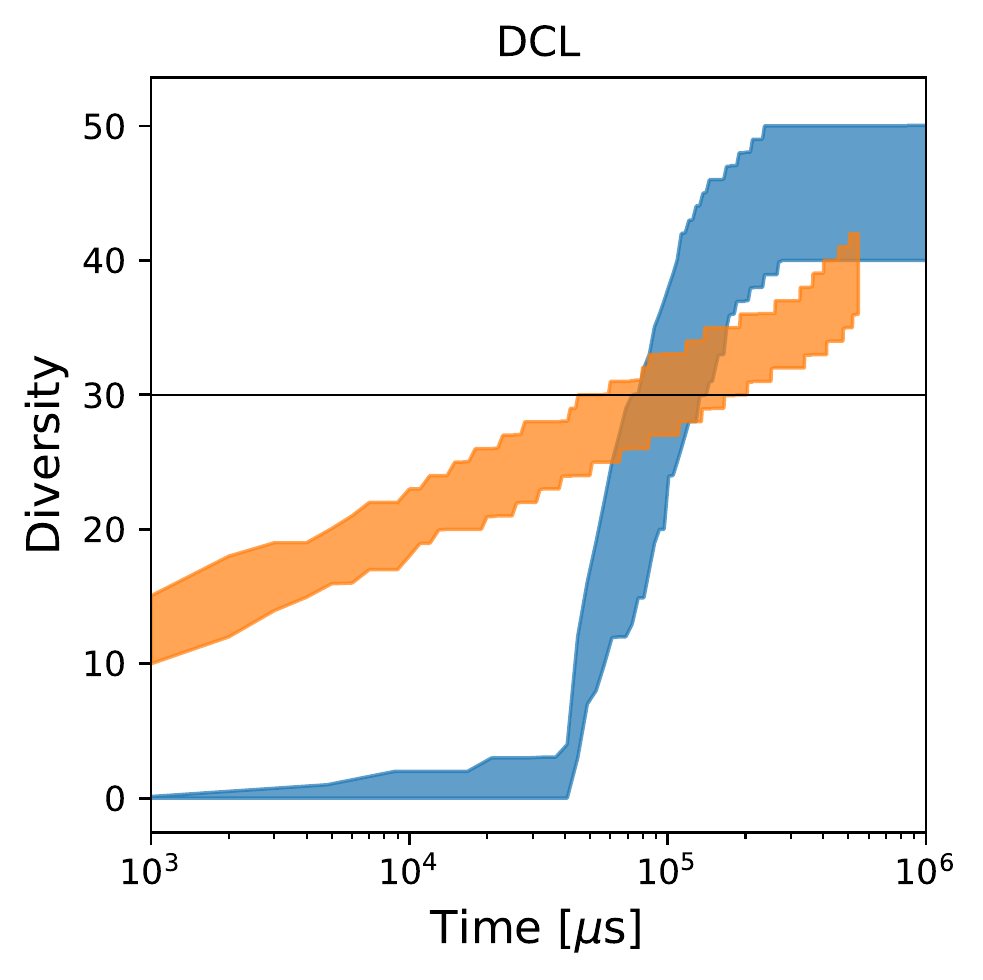}}
    \caption{Diversity VS time for three problems on $C_{12}$, or 1152 variables. The bands show the region between $5^{\rm th}$ and $95^{\rm th}$ percentiles for the lower bound on diversity. Horizontal black lines show the diversity target calculated from 50k samples from the QPU.
    We can readily see the different behavior between QA and PT+ICM. PT+ICM spends a considerable amount of time for burn in before settling in good quality basins and starting to explore good quality and diverse solutions. After that, the diversity of the samples quickly raises until a plateau whose height depends, as we explain in the main text, on the number of independent instances of PT+ICM. Adding PT+ICM instances moves the blue bands up and on the right as more time is spent burning in (because we consider sequential copies of PT+ICM). Conversely, QA does not need to burn in and the diversity of the samples returned steadily increases. From the three problem instances we see that QA is competitive for all the problems and outperform PT+ICM for AC3 problems. When short times and lower target diversities are needed, QA clearly outperforms the classical solver. For longer times, or larger target diversities, and for RAN1 and DCL inputs, PT+ICM is able to outperform QA.} 
    \label{fig:increase_diversity_effect}
\end{figure*} 

\begin{figure*}[tbh]
    \centering
    \subfigure{
    \includegraphics[width=.95\textwidth]{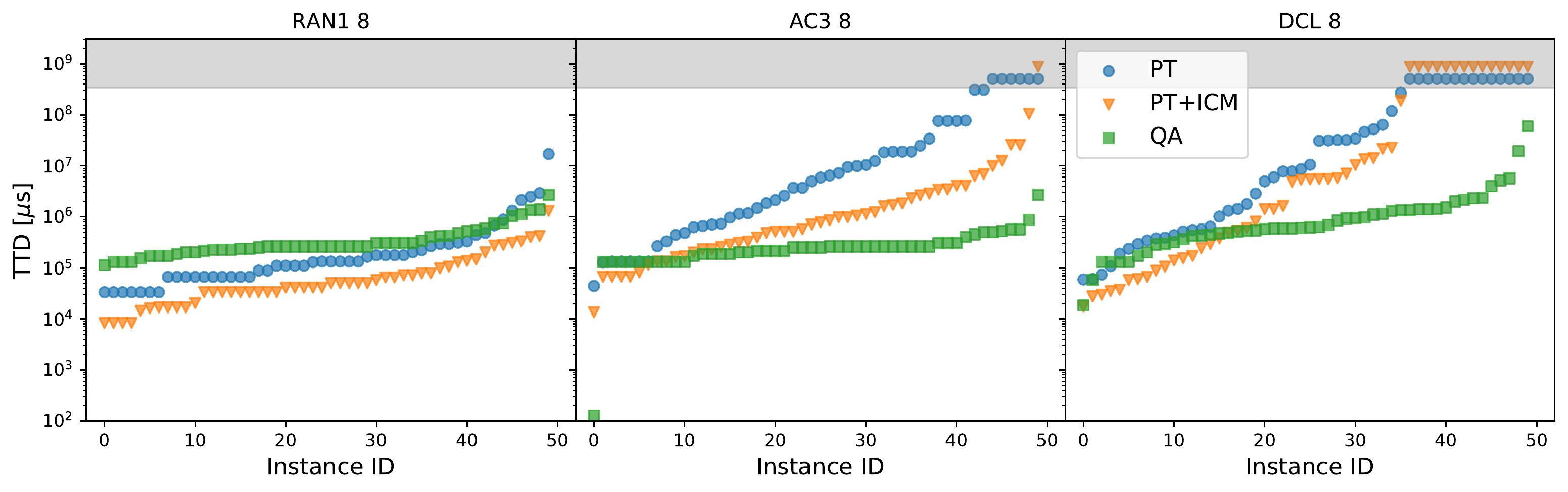}} \\
    \subfigure{
    \includegraphics[width=.95\textwidth]{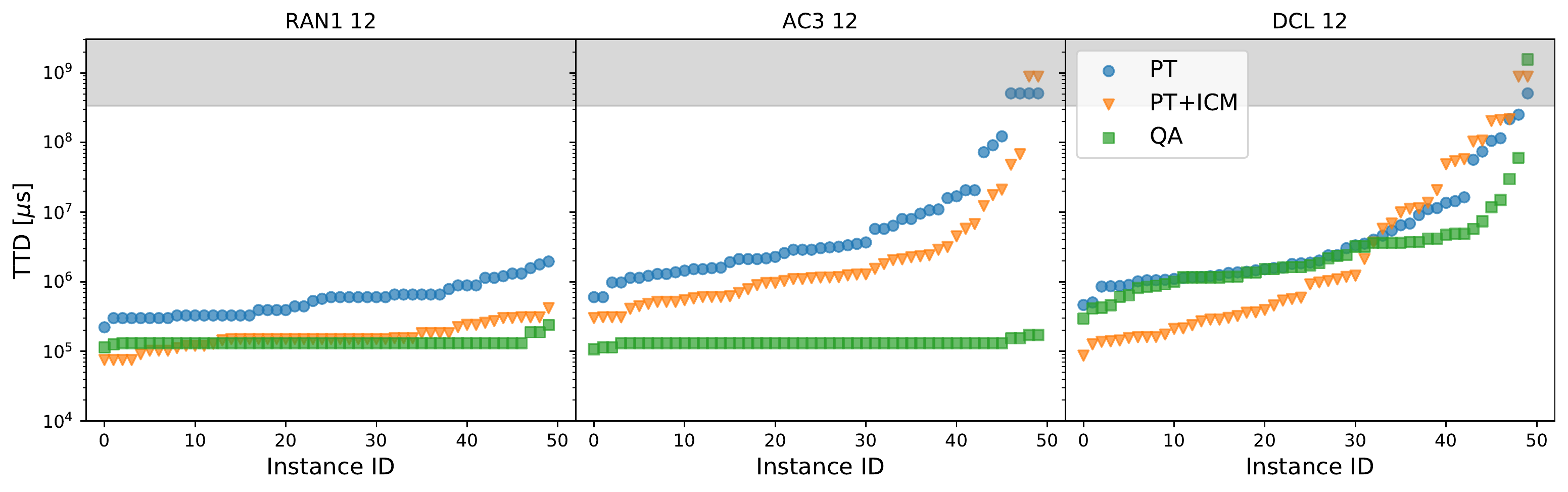}} \\
    \subfigure{
    \includegraphics[width=.95\textwidth]{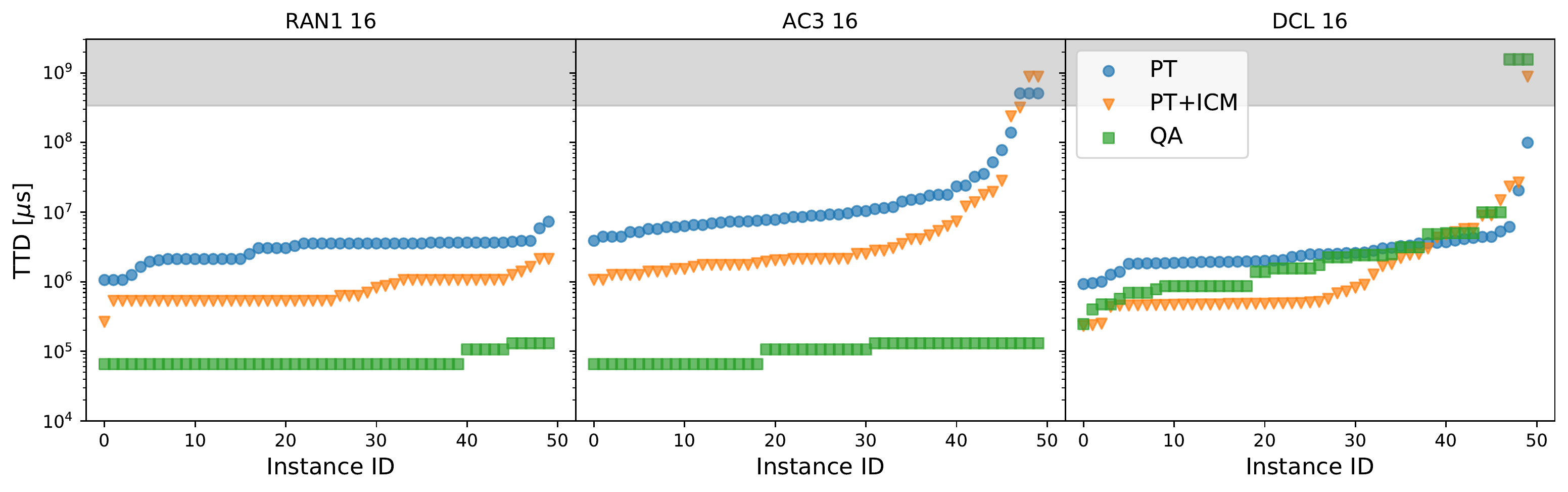}}
    \caption{Comparison of the time-to-diversity (TTD) for parallel tempering (PT), PT with iso-energetic cluster moves (PT+ICM) and quantum annealing (QA) for problem built on a Chimera graph with size $L=8$ (top plots) $L=12$ (center plots) and $L=16$ (bottom plots). The instances are ordered according to increasing TTD. Instances in the grey areas mean that the algorithm reached the wall-time and it was not possible to calculate the exact TTD. On RAN1 problems we see that although the classical solvers outperforms QA at small scale, at large scales QA wins over the classical solvers. An even larger advantage of QA over classical solvers happens for AC3 inputs. For DCL problems, the situation is somewhat different and PT+ICM has a slight advantage over QA.}
    \label{fig:ttd_cdf}
\end{figure*}

\begin{figure*}[tbh]
    \centering
    \subfigure{
    \includegraphics[width=.32\textwidth]{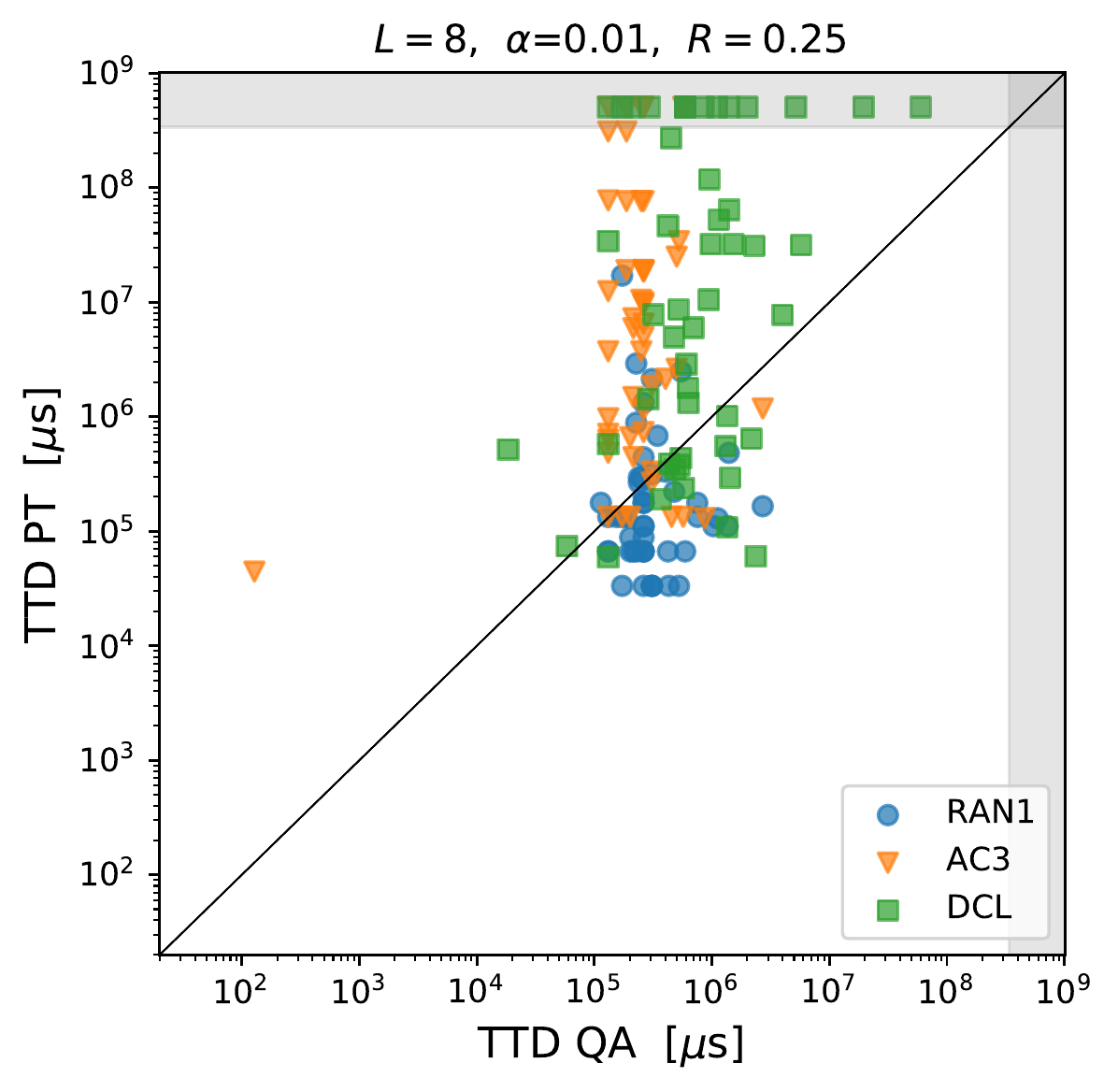}} 
     \subfigure{
     \includegraphics[width=.32\textwidth]{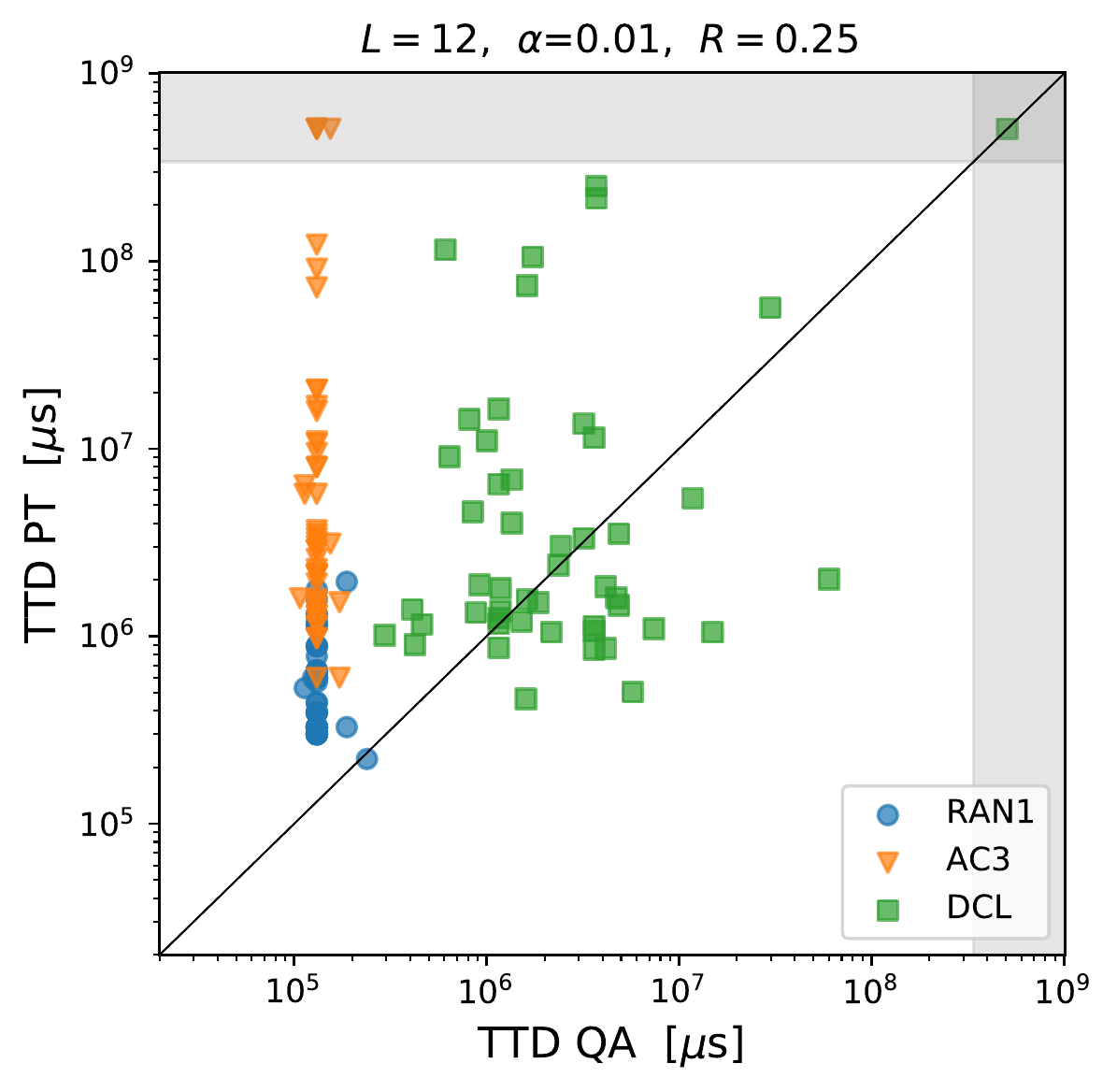}} 
    \subfigure{
    \includegraphics[width=.32\textwidth]{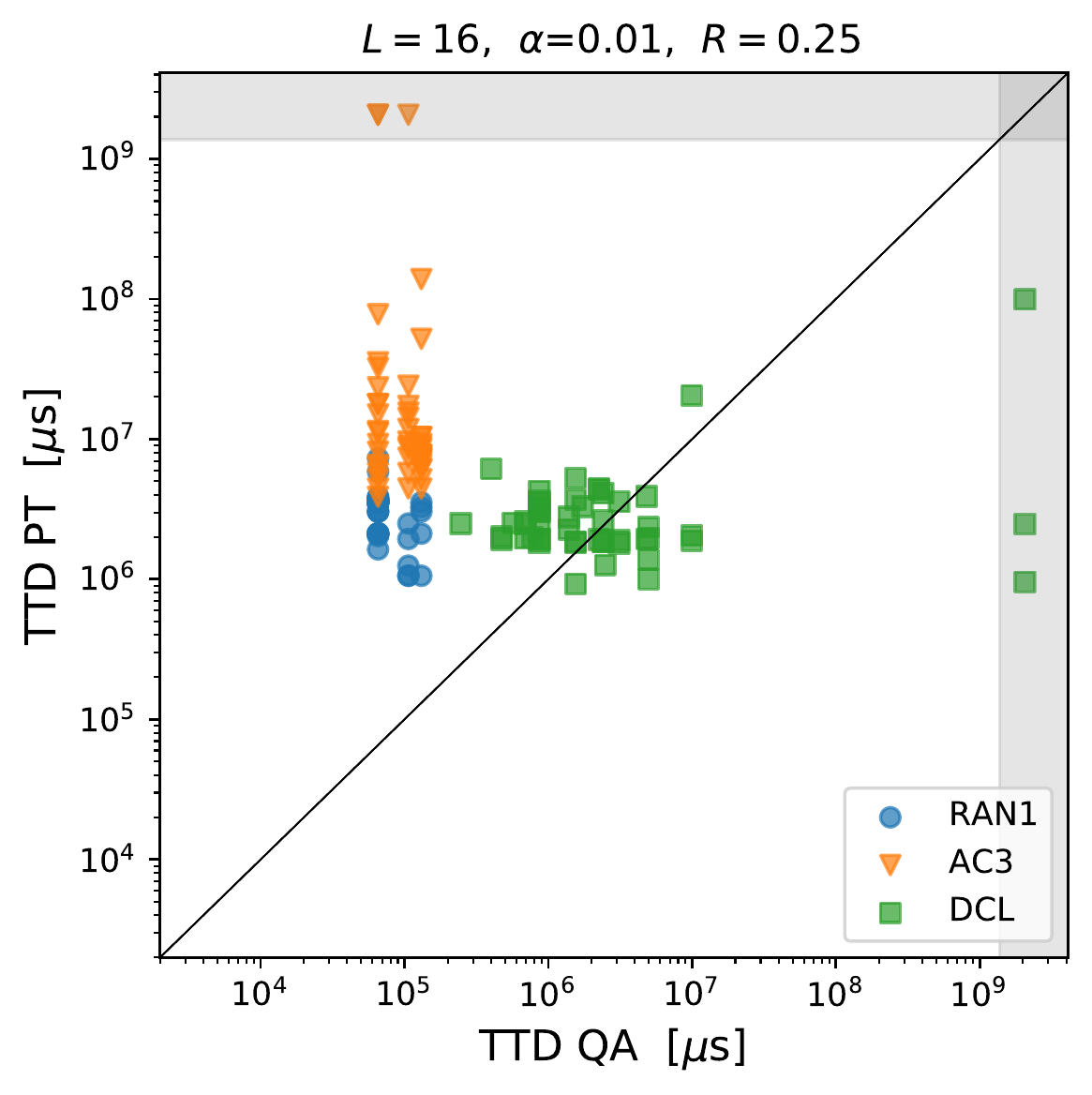}} \\
    \subfigure{
    \includegraphics[width=.32\textwidth]{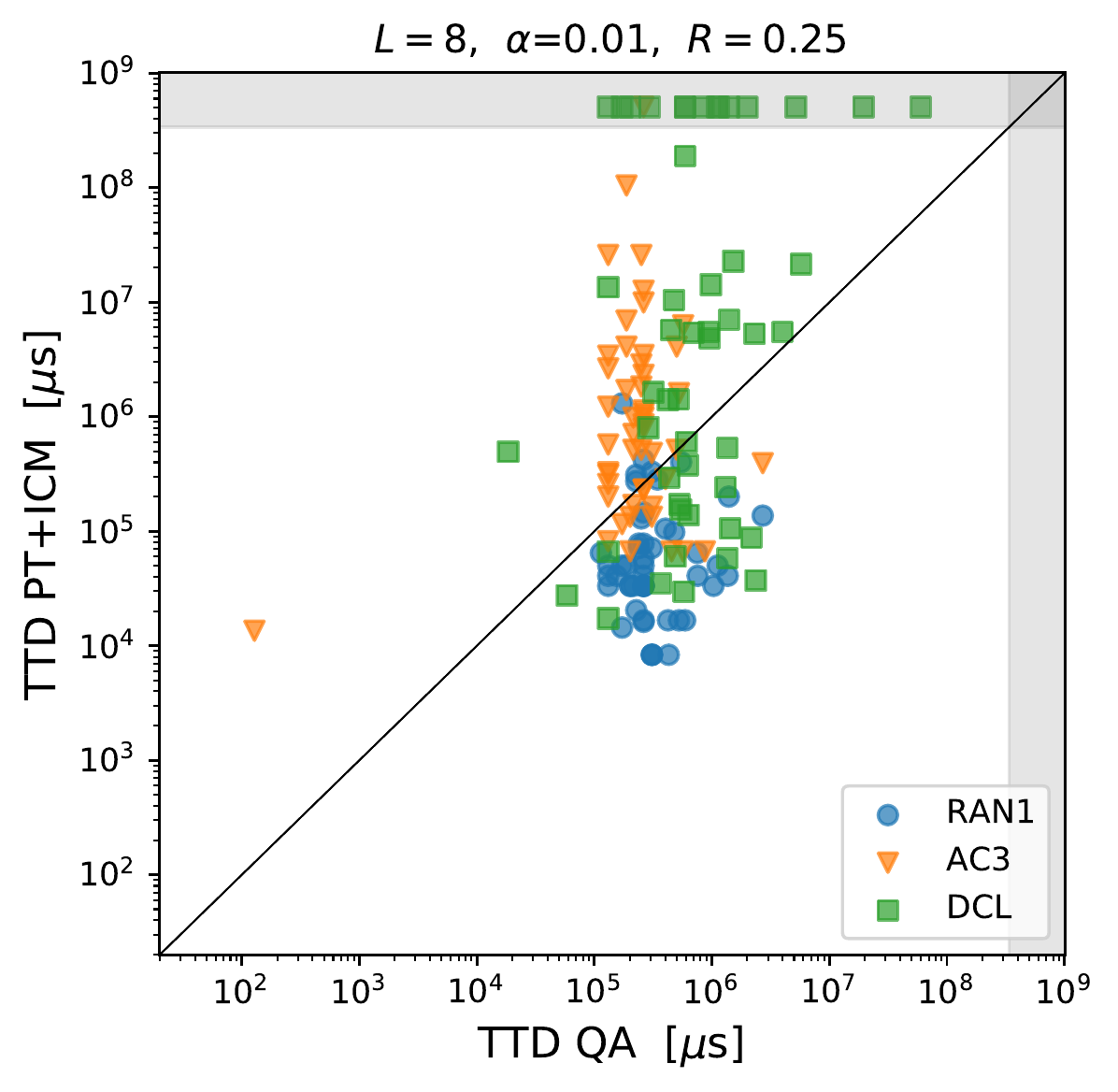}} 
    \subfigure{
    \includegraphics[width=.32\textwidth]{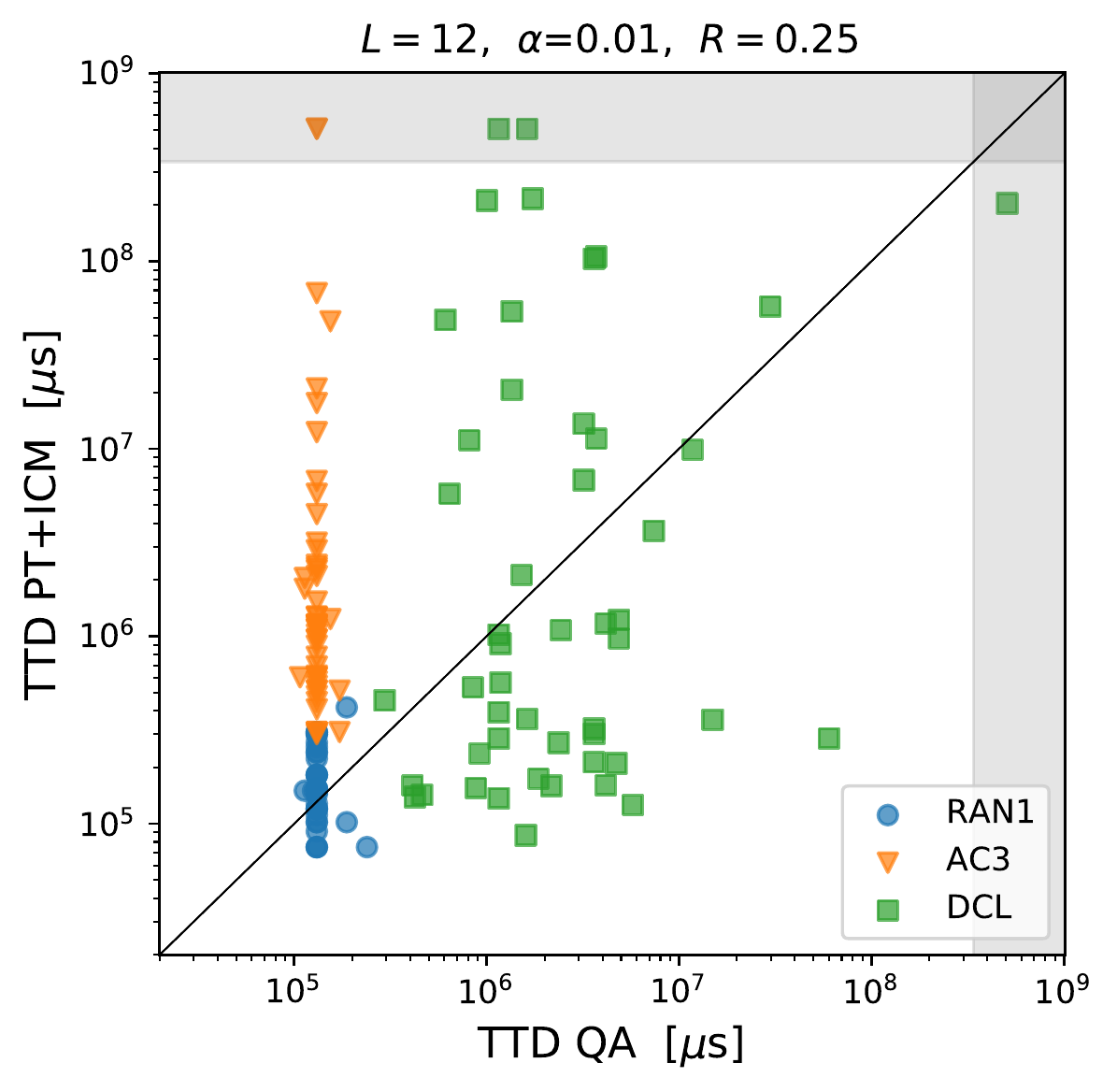}} 
    \subfigure{
    \includegraphics[width=.32\textwidth]{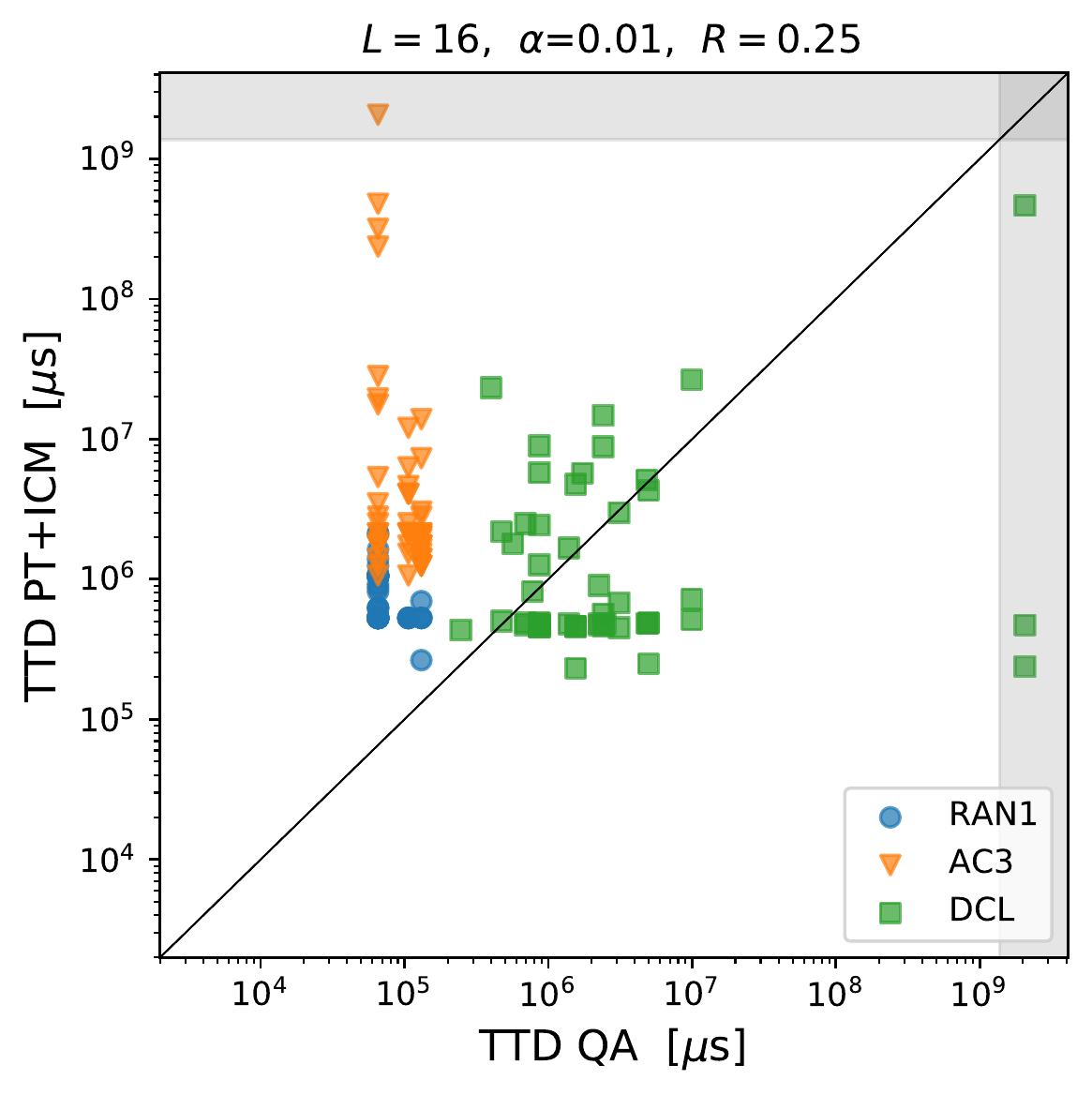}} 
    \caption{Instance by instance comparison of the Time-To-Diversity (TTD) between the the D Wave 2000Q solver and PT (upper plots) and PT+ICM (lower plots), on the three problem instances RAN1, AC3 and DCL on a Chimera size $L=8$ (left plots) $L=12$ (center plots) and $L=16$ (right plots). Points in the grey regions mean that the target diversity was not reached within wall time. A point in the upper left triangle means that, on that particular instance,  QA is winning over the respective classical solver.}
    \label{fig:ttd_instance_comparison}
\end{figure*}

\subsection{Time Measurements}
\label{sec:time_measurements}
In order to compare TTD between the solvers, we made the following assumptions. For QA, we ignore the programming time of the D-Wave 2000Q chip and consider only the annealing time. For the classical algorithms, we ignore the time spent for setting the temperatures, and assume single threaded CPU that can perform a single spin update in 1 ns, so that the elapsed time is given by measuring the total number of spin updates performed. 
When referring to wall-times for classical solvers however, we report actual CPU times on our machines, taken only by the solver.
It is worth noticing that computing  diversity takes much longer than all solver runtimes. 

\subsection{Experimental Setup}
Our experiments consist of four parts: minimum and maximum energy search, target diversity determination, solvers optimization and performance comparison. Below we describe each part.
\vspace{1mm}

\paragraph*{\it Minimum and maximum energy search.} We approximate the ground state $E_0$ with the minimum energy $E_{\rm min}$ among a set of up to  $2\times10^6$ candidate solutions from the PT+ICM and QA solvers. For the maximum energy, we used the fact that for bipartite graphs with zero field $h_i=0$, $E_{\rm max} = - E_{0}$.
\vspace{1mm}

\paragraph*{Target Diversity Determination.} 
As already mentioned, to estimate the target diversity we use the method suggested in Sect.~\ref{sec:practical_target_diversity} and calculate the diversity of 50k samples from the QPU. The diversity is taken as the maximum of 100 independent calculations over the 50k samples.
\vspace{1mm}

\paragraph*{Solvers optimization.} 
We optimized the solvers by a grid search over their hyper-parameter space. For all solvers we chose a subset of the generated problems, and computed TTD for each parameter choice. 
 PT and PT+ICM we tuned the exchange rate to set the temperatures and the number of independent runs.
We collect the samples from every replica in each PT initialization after a fixed number of sweeps. We then calculate diversity using all samples and compare with the target diversity. We continue this process until the target diversity is reached. We found that for both PT and PT+ICM the optimal exchange rate is around $\sim 0.4-0.5$, similarly to the usually adopted exchange rate used for optimization, while the number of independent runs is $\sim 25$ for PT and $\sim 10$ for PT+ICM. For the latter however, we note that the ICM moves include an extra factor of 2 in the number of replicas. 

For QA we optimized the annealing time as well as the number of spin-reversal transformations.
We found that the lowest possible annealing time of 1 $\mu$s minimized the ${\rm TTD}$, while increasing the number of spin-reversal transformations helps to increase the diversity.

Details of the optimization are shown in  App.~\ref{sec:solver_optimization}.

\subsection{Solvers Behavior}
Before presenting the results of our analysis, we briefly discuss behaviors of the QPU and PT+ICM on the three types of problem (PT behaves similarly as PT+ICM and as we show in the next section, performs slightly worse than PT+ICM). In Fig.~\ref{fig:increase_diversity_effect} we plot the diversity reached by a solver over time for three problems - one for each input class - on the $C_{12}$ graph. The bands show the $5^{\rm th}$ and $95^{\rm th}$ percentiles of the distribution of diversity, while the black solid lines are the diversity threshold set by the 50k QPU samples.
Here we can see the fundamental difference between the two types of solvers.  PT+ICM has low diversity during the burn-in time, then quickly increases when each instance of PT access low energy configurations and eventually saturates, possibly because it remains stuck in a limited region. On the contrary, QA is able to produce a set of diverse solutions very quickly and the diversity of the sample set increases steadily with time. More importantly, during the times explored in this analysis, QA does not saturate meaning that the QPU does not get stuck in a certain region in the solution space. 

From Fig.~\ref{fig:increase_diversity_effect} it is also possible to qualitatively inspect what happens when the target diversity is changed.
By increasing the number of samples of the D Wave solver, we effectively raise the horizontal lines for each problem, the exact amount depending on the steepness of the orange bands. In all three problems, increasing the target diversity above the saturation level, the classical solver with the current (optimized) setup will not be able to reach the target diversity obtained by the QPU within a reasonable time. This however may be fixed by another choice of hyper-parameters, for example by increasing the number of copies per temperature, but that will also increase TTD as the classical solver will need more total time to converge.\footnote{With this we mean that although each copy spends approximately the same time to converge, having more copies considerably increase the cumulative time spent.}  Moreover, one also needs to take into account the time required to find a new set of hyper-parameters for each setting.

For the specific RAN1 instance showed here, an increase of target diversity of about 15\% - obtained using approximately 100k samples from the QPU - would degrade the relative advantage of the QPU over PT+ICM. For the AC3 problem shown in the center panel, QPU has an advantage at all the times shown. Increasing or decreasing the samples used to calculate the target diversity would increase the advantage. For the DCL problem, increasing the diversity threshold would make PT+ICM win over QA until the plateau at $t \gtrsim 3\times 10^5$ $\mu$s. Beyond that limit, as already mentioned, one should re-optimize the hyper-parameters of the solvers and investigate the solution again. 

\subsection{Performance Comparison} 
Here we show the performance comparison of the solvers. For each solver and for each problem instance we performed 100 independent experiments, each of them consisting of measuring diversity over computation time,  as described in Sect.~\ref{sec:time_measurements}. For the classical solvers we imposed a CPU wall time of 5 minutes, while for the QPU we imposed a maximum number of analyzed samples of $10^6$. The main results are shown in Figs.~\ref{fig:ttd_cdf} and~\ref{fig:ttd_instance_comparison}. In the former we show the results for the three solvers, ordered by increasing TTD from left to right. The left panels show the results for the RAN1 input instances, the center are the AC3 instances, and the right panels are the DCL inputs. Moving vertically we change the size of the Chimera graph. For the AC3 instances we see that the QPU performs better than the other classical solvers, with QPU relative performances improving with larger sizes. This can also be seen in Fig.~\ref{fig:ttd_instance_comparison}, where we show instance by instance comparison between the solvers. On the top panels we show the comparison between PT and QA, while on the lower panels we compare the QPU with PT+ICM. Moving horizontally we increase the size of the Chimera graph. 

In both figures, we see that for AC3 instances (yellow triangles) the QPU is faster than the two classical solvers and how for increasing sizes the instance by instance comparisons ``move up'' in the plots. Likewise for RAN1 instances, the QPU has better performances, although on $C_8$ the classical solvers have comparable, if not better performances. On C8 in fact, as can be seen in the lower left panel on Fig.~\ref{fig:ttd_cdf}, PT+ICM is  faster than QA. We note that for RAN1 and AC3, the TTD for QPU is almost constant among the instances. 
This is probably due to the fact that the target diversity is set by the QPU. 
For DCL problems the behaviour is different and the performance of the QPU is comparable with that of PT and PT+ICM and does not get better with increasing size. This could be related to the fact that DCL parameters require more precision. 

\subsection{Decreasing the approximation ratio}
\label{sec:small_alpha}

\begin{figure*}
    \centering
    \subfigure{
    \includegraphics[width=.75\textwidth]{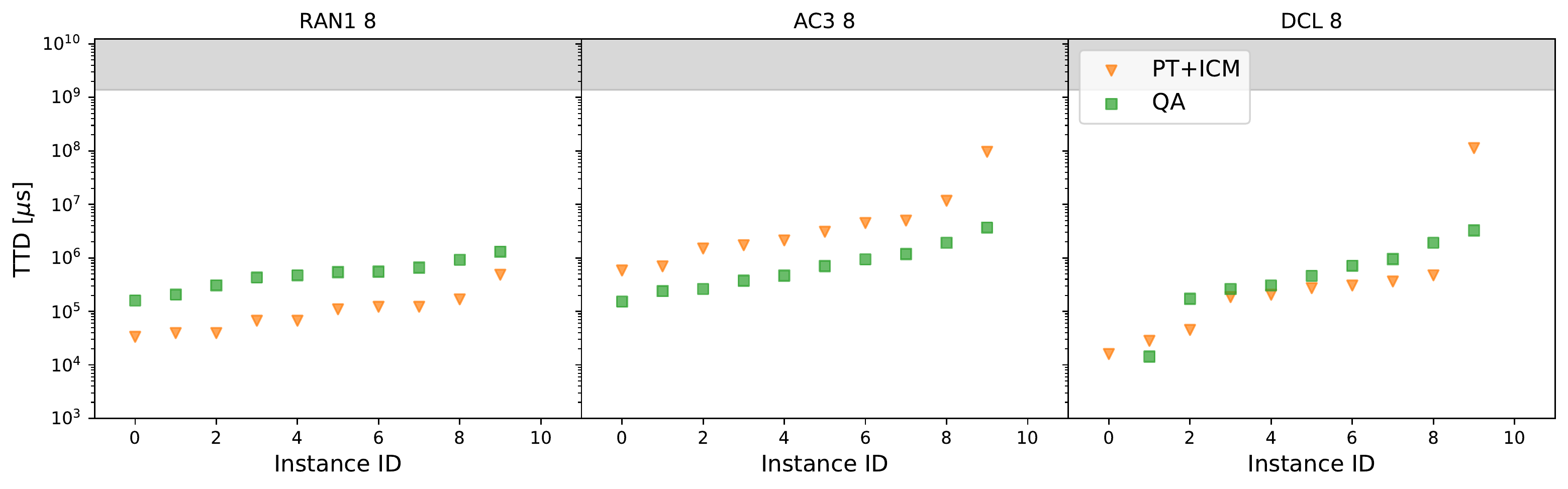}} \\
    \subfigure{
    \includegraphics[width=.75\textwidth]{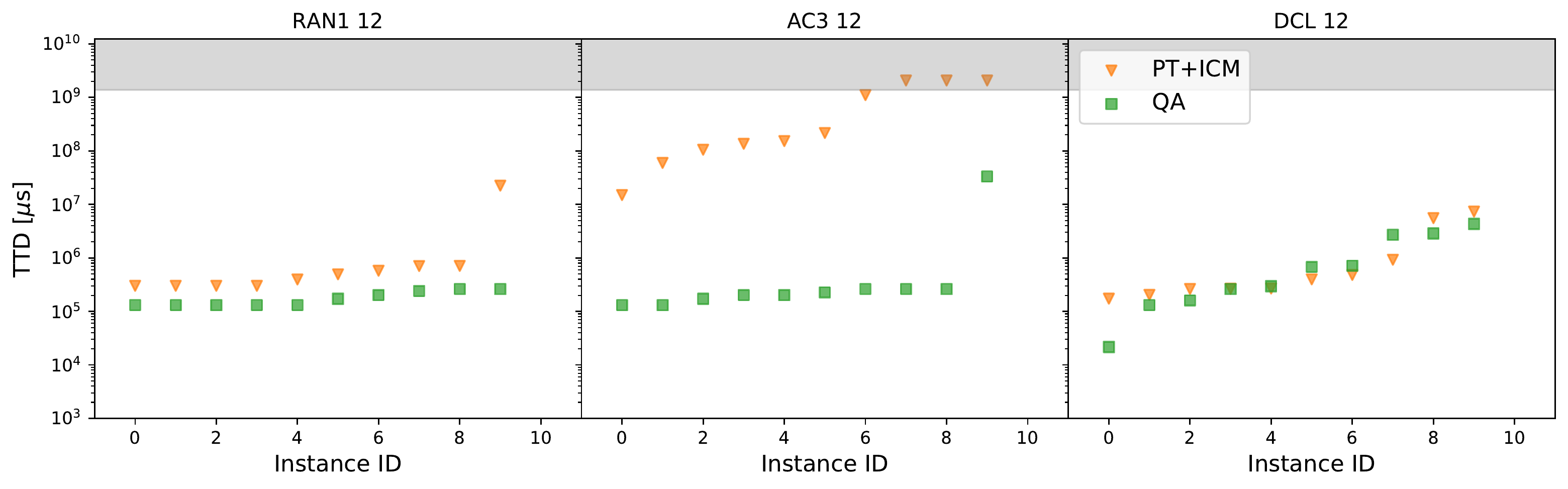}} \\
    \subfigure{
    \includegraphics[width=.75\textwidth]{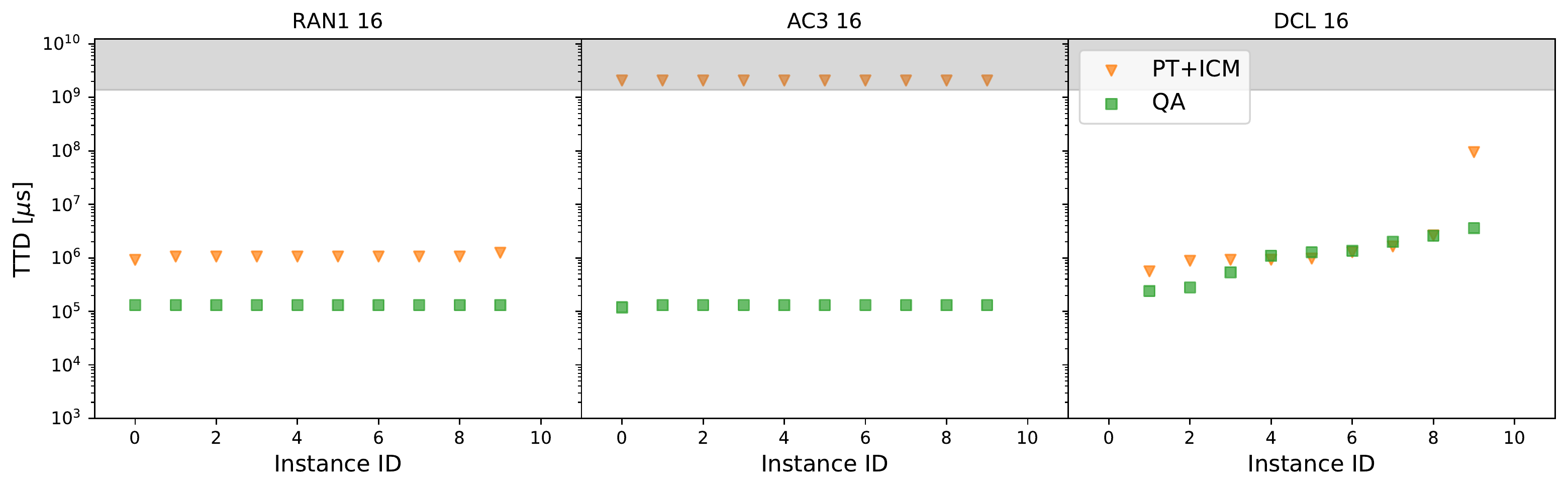}}
    \caption{Comparison of TTD with approximation ratio $\alpha=0.005$ for PT+ICM and QA for problems on $C_8$ (upper panels), $C_12$ (mid panels) and $C_16$ (lower panels). From left to right the inputs are RAN1, AC3 and DCL. Here the situation is similar to Fig.~\ref{fig:ttd_cdf}: for RAN1 inputs, QA performance relative to PT+ICM improves as the scale of the problem is increased. For AC3 QA outperforms the classical solver at all scales and for DCL problems, interestingly, the performances are on par at all scales.}
    \label{fig:cdf_lower_alpha}
\end{figure*}
Here we briefly discuss what happens when a lower approximation ratio is required. For lower $\alpha$, TTD will be dominated by the TTT metric  \cite{2015arXiv150805087K}. We  focused here on $\alpha=0.005$ and to limit computational time,  we selected only a small batch of the problems above. Also, because PT+ICM performed consistently better than PT, we only considered the former. We re-optimzed the solvers and found that for PT+ICM  the optimal exchange rate dropped to $0.2$ while the number of copies per temperature remained the same, 10. However, in this case the optimal region in the parameter space is larger, as can be seen in Fig.~\ref{fig:pt_optimization_lower_alpha}. For the QPU the optimal parameters did not change.

The results for this approximation ratio are shown in Fig.~\ref{fig:cdf_lower_alpha} and they are similar to the case of $\alpha=0.01$. For AC3 problems, TTD for QA is consistently smaller than for PT+ICM. In the case of RAN1 problems, PT+ICM performs better for smaller problems, but TTD scales in favour of the QPU for larger sizes. For DCL inputs, the performances between the two solvers are comparable for all sizes.
Due to calibration errors, noise and finite temperature, we expect the performance of the QPU do degrade as we decrease the approximation ratio to lower values. Although $\alpha=0.005$ is already a good target for many applications, future generations of the QPU are likely to have less calibration errors and lower noise, making smaller approximation ratios achievable.


\section{Conclusions}
\label{sec:conclusions}
In this work, we have used a new metric to assess the diversity of the samples returned by quantum annealers. In principle, many definitions of diversity are possible, from clustering-inspired metrics to other metrics like average Hamming distance between a sequence of solutions. Here, the diversity is the maximum independent set of the graph of samples below an approximation ratio $\alpha$, where the edges are defined only between samples within a Hamming radius $R$. We believe that such a diversity metric can have a practical relevance in the context of hybrid genetic algorithms and quality diversity optimization \cite{10.3389/frobt.2016.00040}. Although computing the metric is \#P-hard, we identified an algorithm that can quickly compute its lower bound,  {\it i.e.} it scales linearly with the number of solutions. Similarly to the Time-To-Solution (TTS) metric, we have adopted the notion of  Time-To-Diversity (TTD) as a metric to assess the speed of a heuristic to obtain a target diversity of a problem. 
We then assessed the performance of the solvers to reach the reference diversity for the case $\alpha=0.01$ and $R=0.25$. We observed that for RAN1 and AC3 problems, the D Wave quantum annealer shows a performance advantage over the classical solvers, with the advantage increasing with problem size.  
For the DCL inputs, the QPU performances are comparable with PT+ICM at all sizes. We also considered a case in which the approximation ratio is reduced to $\alpha=0.005$ and found that the results were similar to the case of $\alpha=0.01$. This suggests that, for applications were large diversity of samples is required, a portfolio solver that combines quantum and classical heuristics will win over other solvers.

\acknowledgments
This work was supported by Mitacs through the Mitacs Accelerate program.
The authors would like to thank Jeremy Hilton, Emile Hoskinson, Andrew King, Eric Ladizinsky, Jack Raymond and Trevor Lanting for useful discussions. 


\newpage
\bibliography{references.bib}

\clearpage
\appendix
\widetext
\section{Algorithms for approximated diversity}
\label{sec:approximated_diversity}
Here we provide the details of the algorithms used to compute lower and upper bounds on diversity.

\subsection{Lower bound: left-neighbors-aggregation algorithm}
\label{sec:lna_algorithm}
We begin with the agglomerative algorithm used to compute the lower bound on the diversity.
Given the samples returned by a solver, we only keep the unique fit samples in order of appearance. This defines a \emph{labelled} graph $G= (V,L,E)$ where $L:V \to \{ 1, \dots , m\}$ is the \emph{labelling} that associates each node $v\in V$ with an integer number, $L(v) \in \{1, \dots, m\}$ - in our case, given by the order in which the sampler is returned by the solver.
Given a vertex $v$, we also define its neighbours set N$(v)$ as the set of nodes connected to $v$ through an edge $e \in E$. We also define the left (right) neighbours LN$(v)$ (RN$(u)$) as the subset of the neighbours $u$ of $v$ such that $L(u) < L(v)$ ($L(u)>L(v)$). The algorithm is illustrated in Alg.~\ref{alg:LNA}.

It is clear that the set $\mathcal{S}_{\alpha}^*$ constructed by the LNA algorithm depends on the specific labelling function, and that it can also be a maximum independent set. The latter property is easy to show. Assuming that the graph has a maximum independent set of $\alpha$ elements, one can just pick a labelling $L$ for which the first $\alpha$ elements are the elements of the maximum independent set, and then the other nodes of the graph. The advantage of using this algorithm to find a lower bound on diversity is that its time-complexity is $\mathcal{O}(m \cdot D^{\rm low}_{R \mathcal{S}_{\alpha}})$ at maximum, making it suitable for our purposes.

To improve the lower bound on diversity given by this algorithm, we could re-shuffle the collected samples.

\begin{algorithm}
\caption{Left-Neighbors-Aggregation}
\label{alg:LNA}
\begin{algorithmic}
\STATE {\bf INPUT}: $\mathcal{S}_{\alpha}$: set of fit, unique samples. $|\mathcal{S}_{\alpha}| = m$. Labelling function $L:\mathcal{S}_{\alpha}\to{1, \dots, m}$
\STATE Set $\mathcal{S}^*_{\alpha} = \left\{ L^{-1}(1) \right\}$
\FORALL{$i=2, \dots, n$}
\STATE Set $v = L^{-1}(i)$ 
\IF{${\rm LN}(v) \cap \mathcal{S}_{\alpha}^* = \varnothing$}
\STATE $\mathcal{S}_{\alpha}^* = \mathcal{S}_{\alpha}^* \cap {v}$
\ENDIF
\ENDFOR
\STATE {\bf OUTPUT}:  $\mathcal{S}_{\alpha}^*$.
\end{algorithmic}
\end{algorithm}

\subsection{Upper bound: greedy coloring algorithm}
\label{sec:greedy_coloring}
Here we describe the \emph{greedy-coloring} algorithm \cite{Kosowski2004ClassicalCO} used to provide the upper bound on diversity. As we did with the agglomerative algorithm in App.~\ref{sec:lna_algorithm}, we keep all the samples returned by a solver that have energy within the threshold $E_{\rm th}$ in order of appearance. This defines the complement graph $\bar{G} = (V,L,\bar{E})$ where $L$ is the labelling function defined in App.~\ref{sec:lna_algorithm}, and $\bar{E}$ is the set of edges defined between vertices (samples) that have Hamming distance greater than $nR$. The greedy-coloring algorithm then follows the ordering given by $L$ and is summarized in Alg.~\ref{alg:greedy_coloring}.

\begin{algorithm}
\caption{Greedy - coloring}
\label{alg:greedy_coloring}
\begin{algorithmic}
\STATE {\bf INPUT}: $\mathcal{S}_{\alpha}$: set of fit, unique samples. $|\mathcal{S}_{\alpha}| = m$. Labelling function $L:\mathcal{S}_{\alpha}\to{1, \dots, m}$
\STATE Set $C_1 = {L^{-1}(1)}$, $c = 1$
\FORALL{$i=2, \dots, n$}
\STATE Set $v = L^{-1}(i)$ 
\STATE Set $j=1$
\STATE vertex $v$ is not assigned
\WHILE{(vertex $v$ is not assigned) $  j \le c$ }
\IF{$C_{j} \cap N(v) = \varnothing$}
\STATE $C_{j} = C_{j} \cup \{v\} $
\STATE $v$ is assigned
\ELSE
\STATE j = j+1
\ENDIF
\ENDWHILE
\IF{$v$ is not assigned}
\STATE $C_{c+1} = \{v\}$
\STATE $c = c+1$
\ENDIF
\ENDFOR
\STATE {\bf OUTPUT}: number of colors $c$.
\end{algorithmic}
\end{algorithm}
 Like the agglomerative algorithm in App.~\ref{sec:lna_algorithm}, the greedy-coloring depends on the labelling $L$. Different strategies to choose the labelling can lead to considerable improvement of performances.
 
 \subsection{Algorithms performances}
 \label{sec:algorithm_performances}
\begin{figure}
     \centering
     \subfigure{
     \includegraphics[width=.32\textwidth]{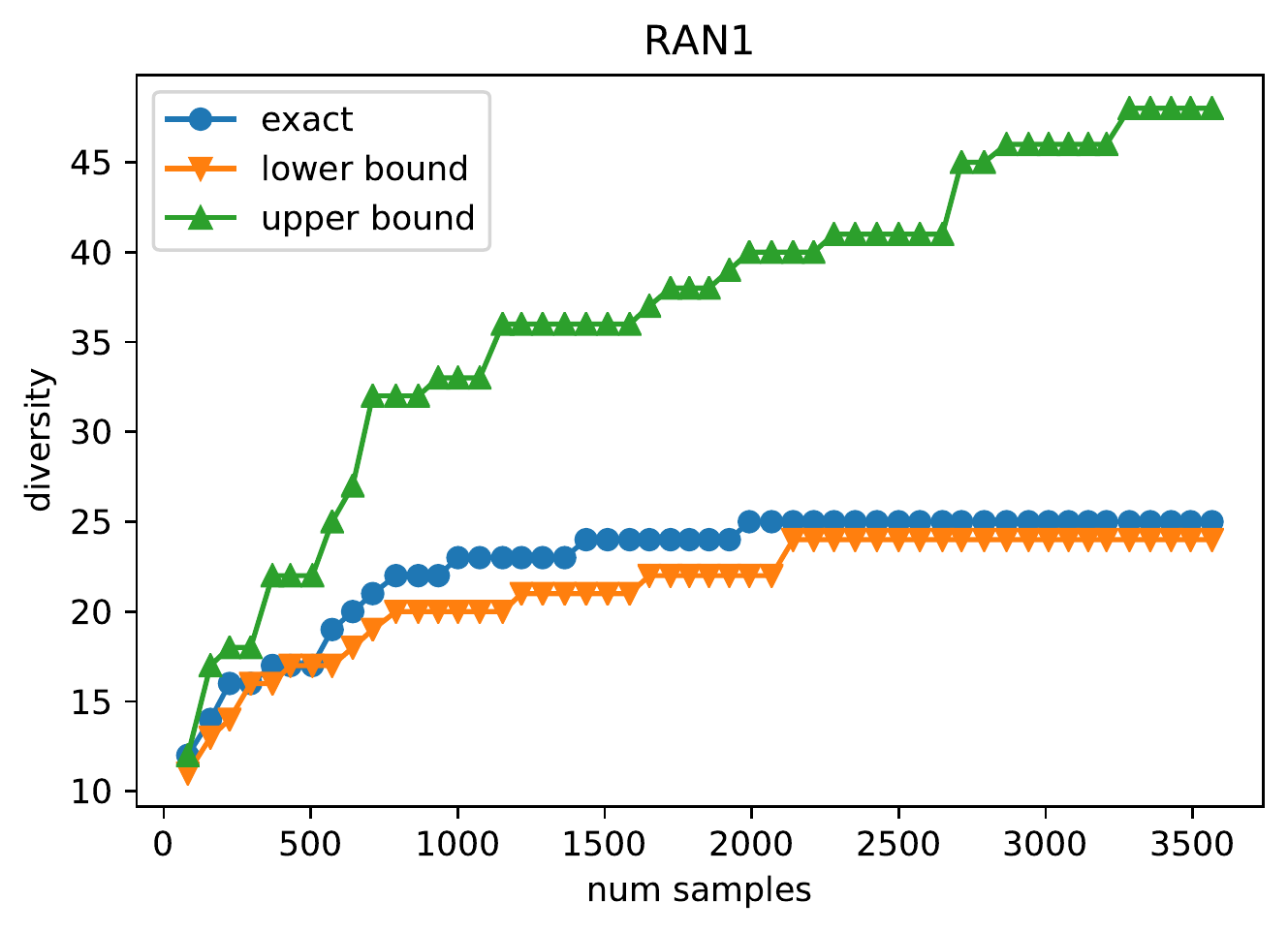}}
     \subfigure{
     \includegraphics[width=.32\textwidth]{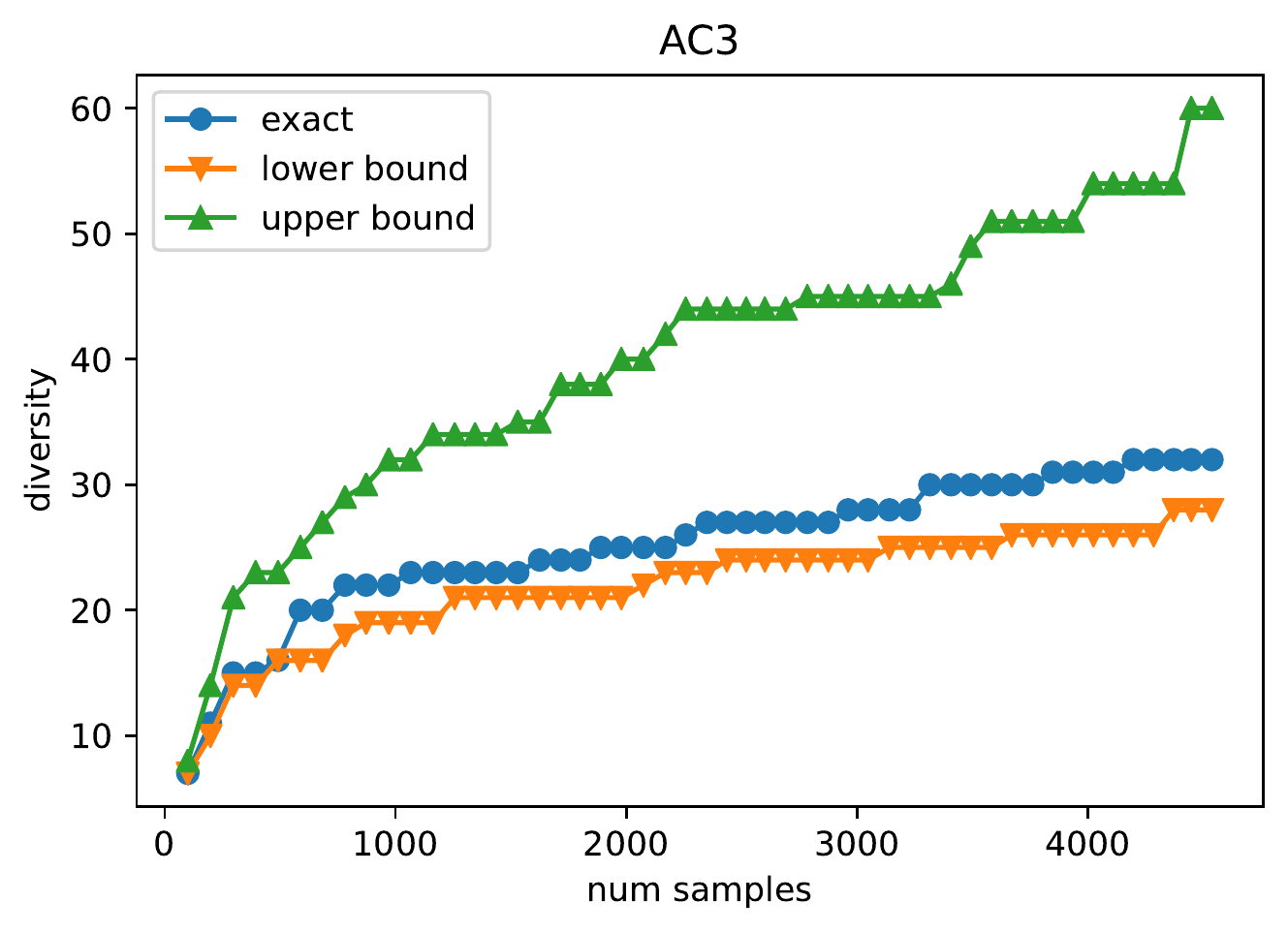}}
     \subfigure{
     \includegraphics[width=.32\textwidth]{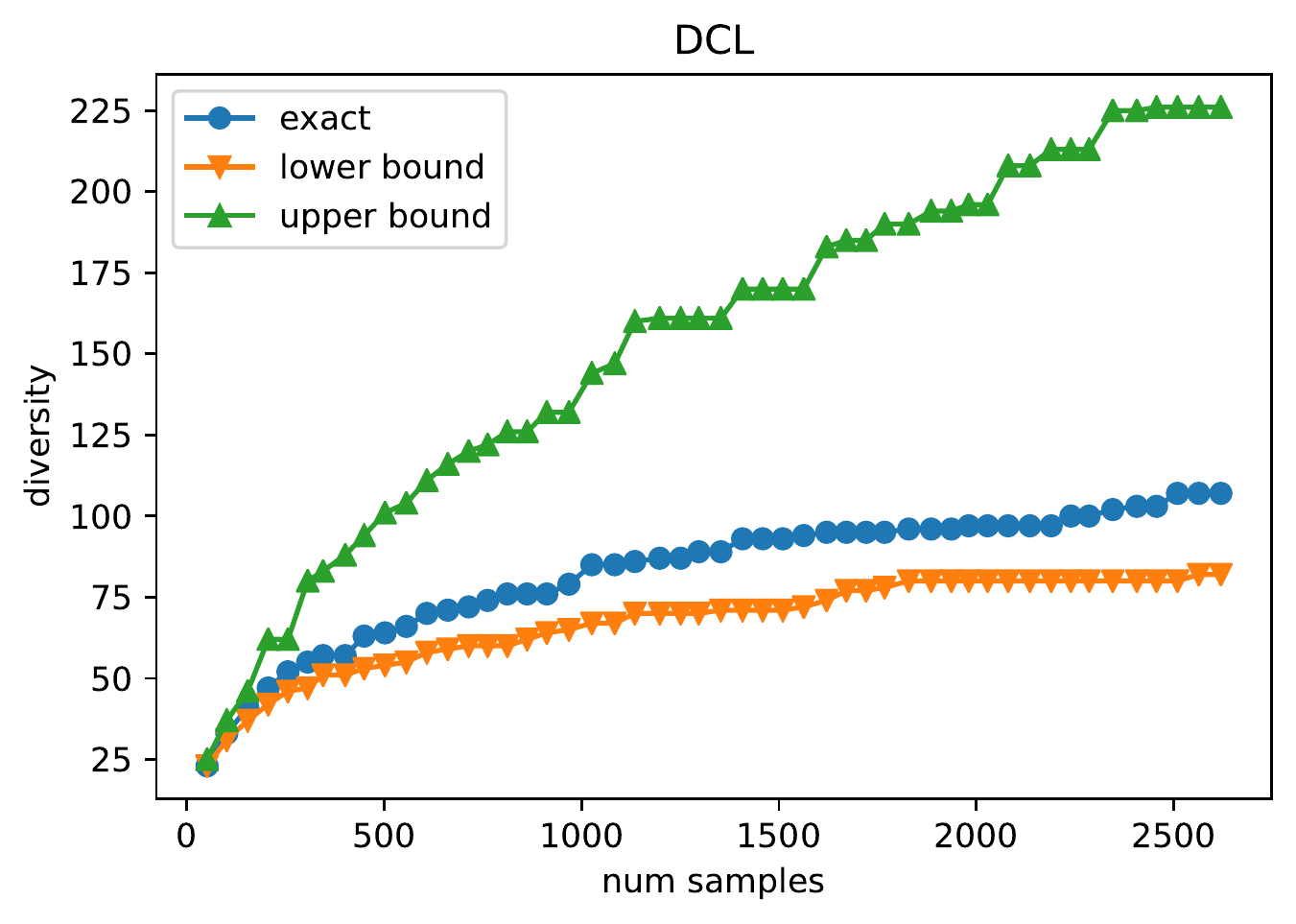}} \\
     \subfigure{
     \includegraphics[width=.32\textwidth]{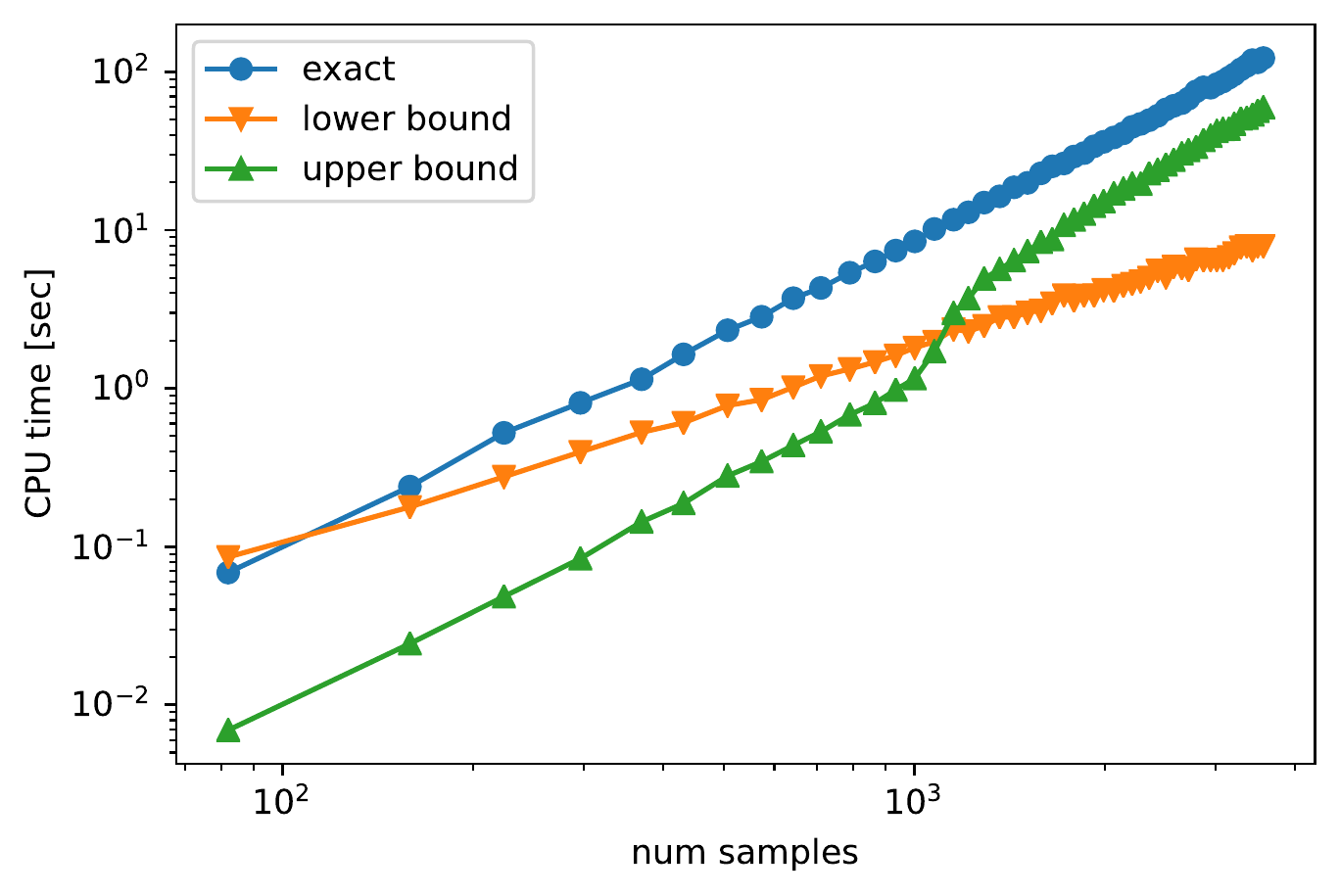}}
      \subfigure{
     \includegraphics[width=.32\textwidth]{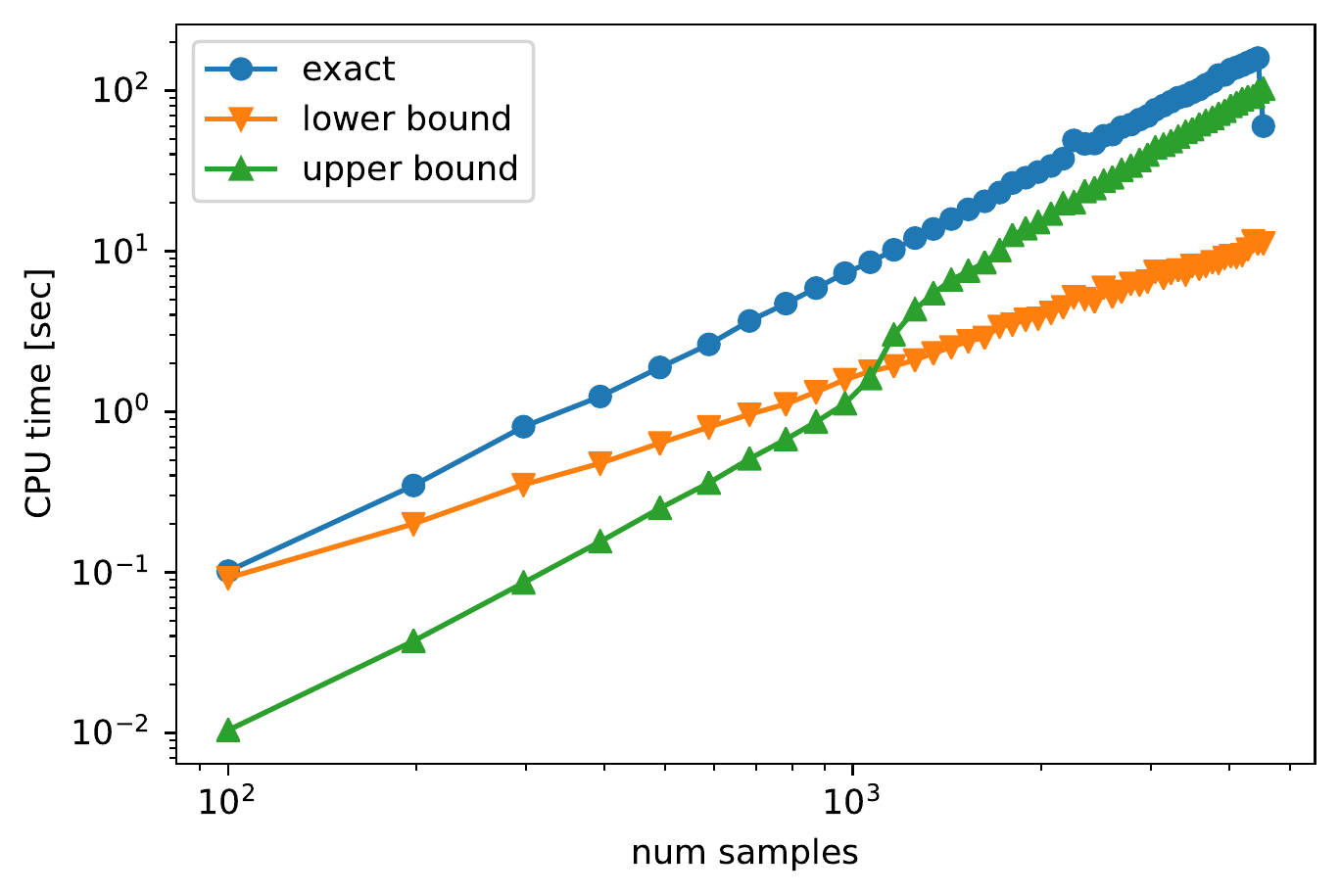}}
      \subfigure{
     \includegraphics[width=.32\textwidth]{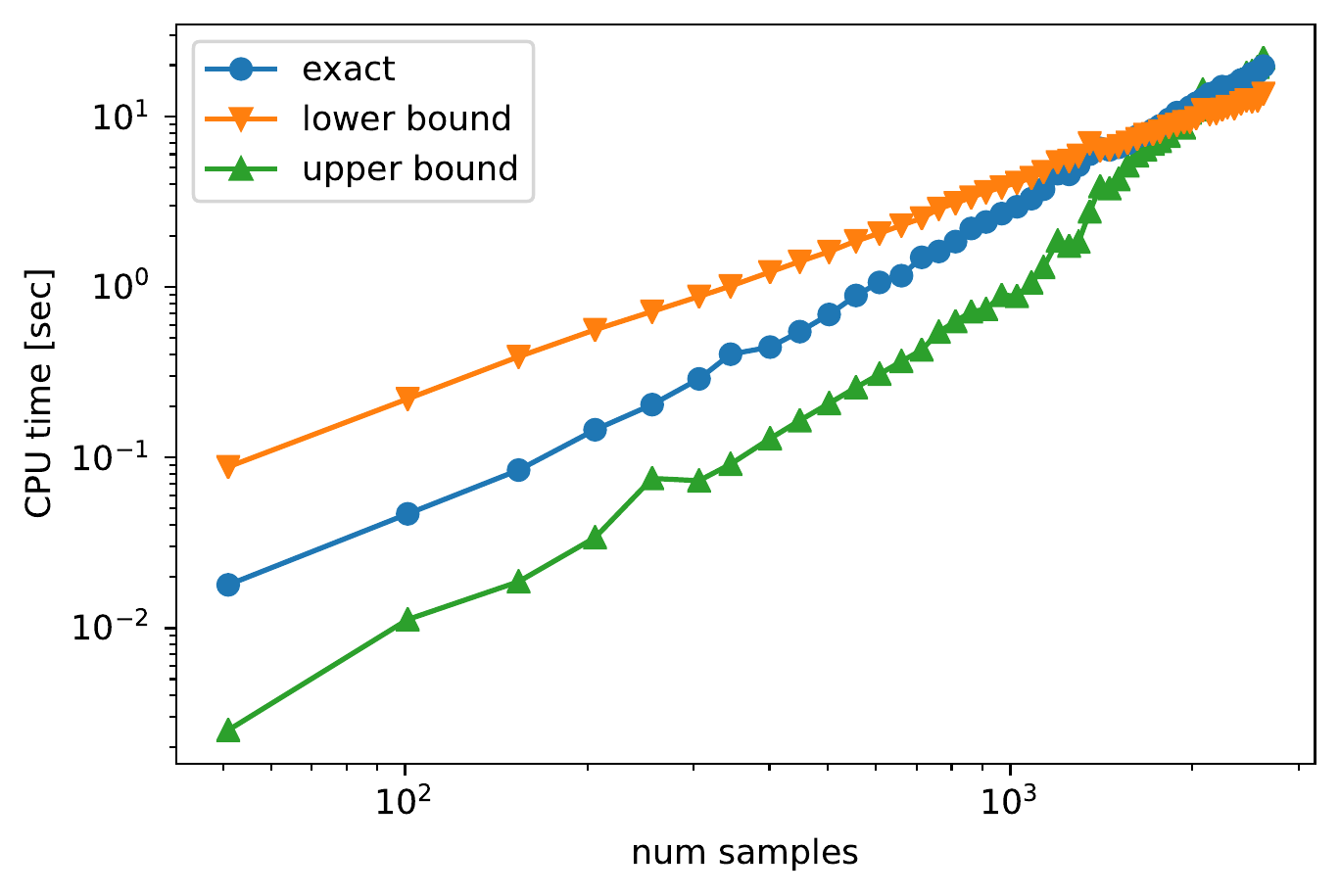}}
     \caption{Comparison between the approximated diversity algorithms LNA and greedy coloring and a tight bound on diversity. The upper panels show the diversity computed by algorithms as a function of samples analyzed for three instances in each class RAN1, AC3 and DCL. The lower panels show the CPU time spent on the computations.}
     \label{fig:algorithm_comparison}
 \end{figure}

 Here we compare the performance of the LNA and greedy coloring algorithms against another tight lower bound on diversity, {\it i.e.}  by mapping the maximum independent set problem to an Ising Hamiltonian \cite{2014FrP.....2....5L} and solving with via simulated annealing (SA). In Fig.~\ref{fig:algorithm_comparison} we show the comparison of the algorithms for one problem in each class analyzed in this paper. The mapping to Ising Hamiltonian is denoted by ``exact'' meaning that we are solving the exact maximum independent set problem with the SA heuristic. To improve the accuracy of the LNA algorithm, we reshuffle the samples 100 times. We can see how with this setting - used also in our analysis in Sect.~\ref{sec:results_and_discussion}  the lower bound is remarkably close to the values obtained by solving the maximum independent set problem. Conversely, the upper bound is not a tight bound and diverges from the exact value rather quickly. Importantly, the computation time for the LNA method has a better scaling compared to the other two methods. This motivates us to adopt the lower bound quantity for our analysis.

\section{Solvers optimization}
\label{sec:solver_optimization}
In this Appendix we describe the solver optimization. 
\subsection{Parallel Tempering}

\begin{figure}
    \centering
    \includegraphics[width=.9\textwidth]{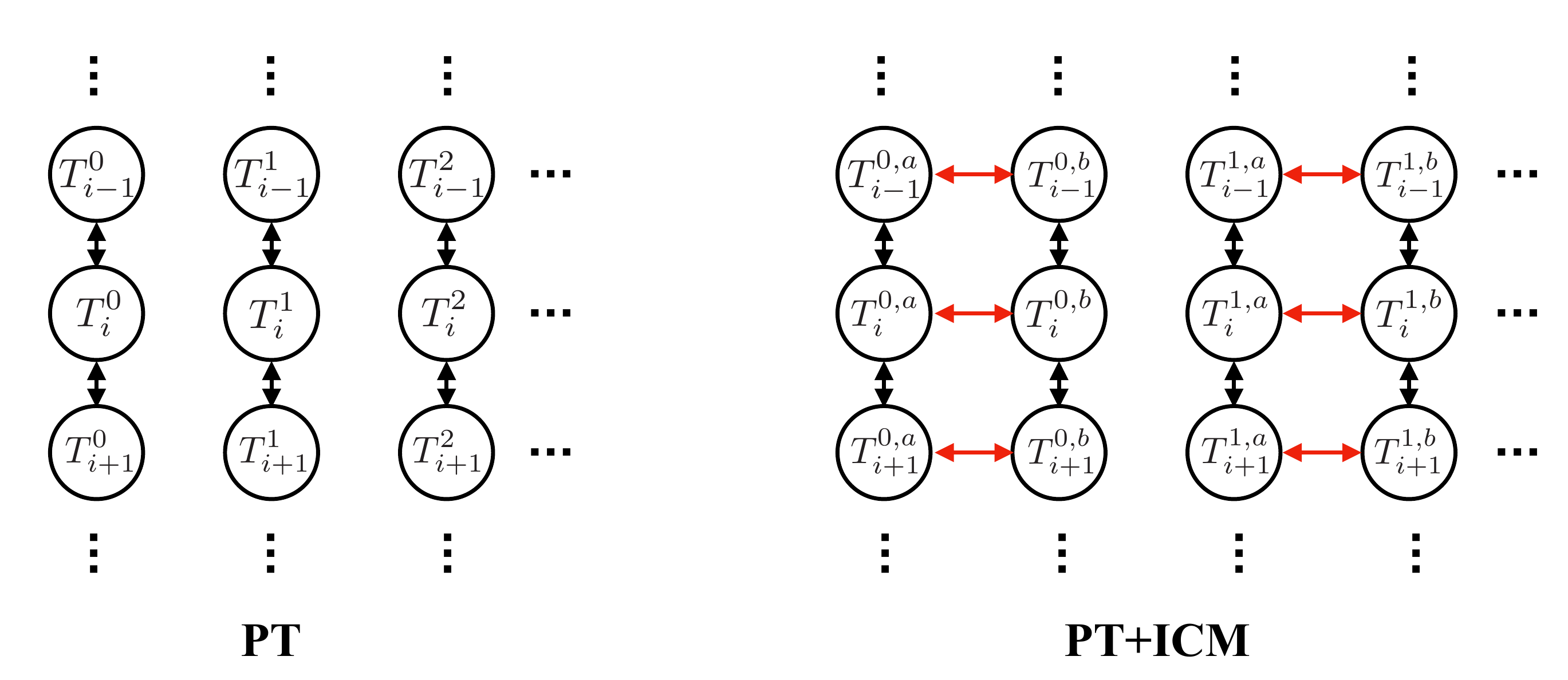}
    \caption{Configuration of PT (left side) and PT+ICM (right side) used in this work. Black vertical arrows denote replica exchange due to thermal fluctuations, while red horizontal arrows denote the Houdayer iso-energetic cluster moves, performed between replicas at the same temperature.}
    \label{fig:pt_config}
\end{figure}

\begin{figure}
    \centering
    \subfigure{
    \includegraphics[width=.3\textwidth]{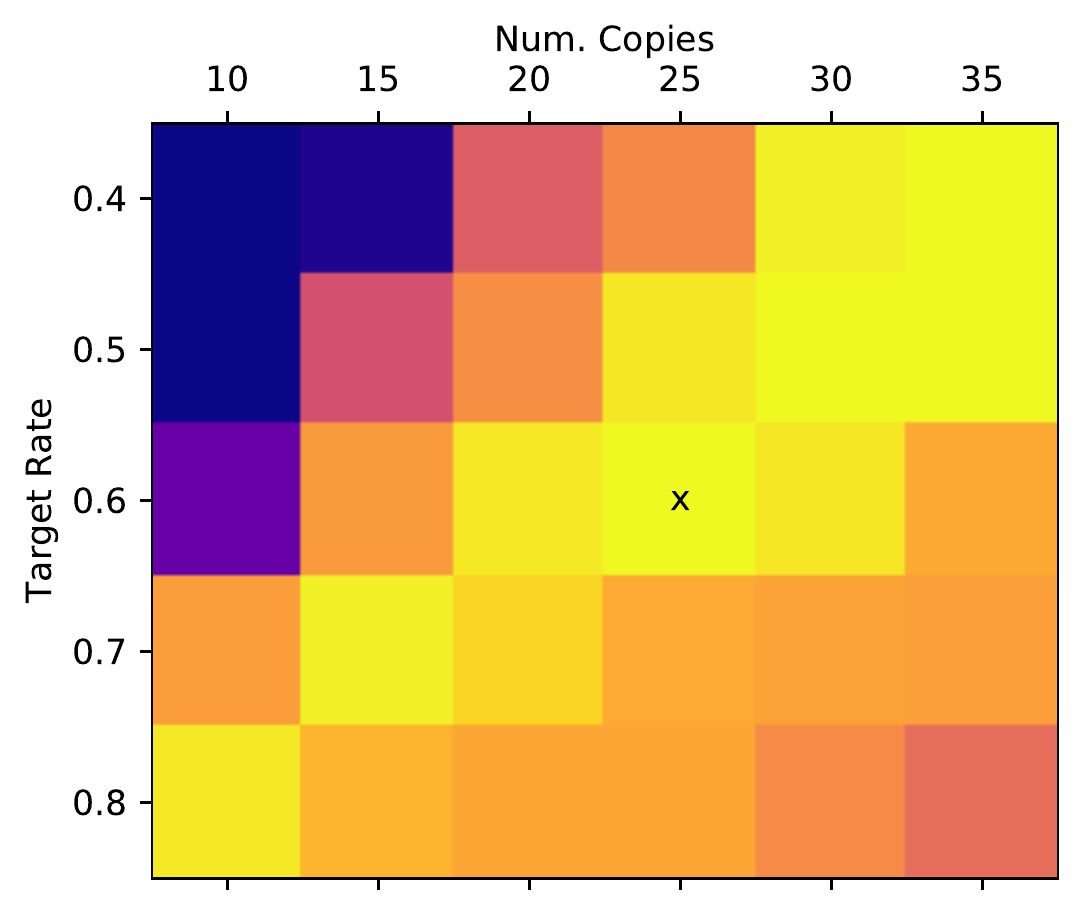}}\,
    \subfigure{
    \includegraphics[width=.3\textwidth]{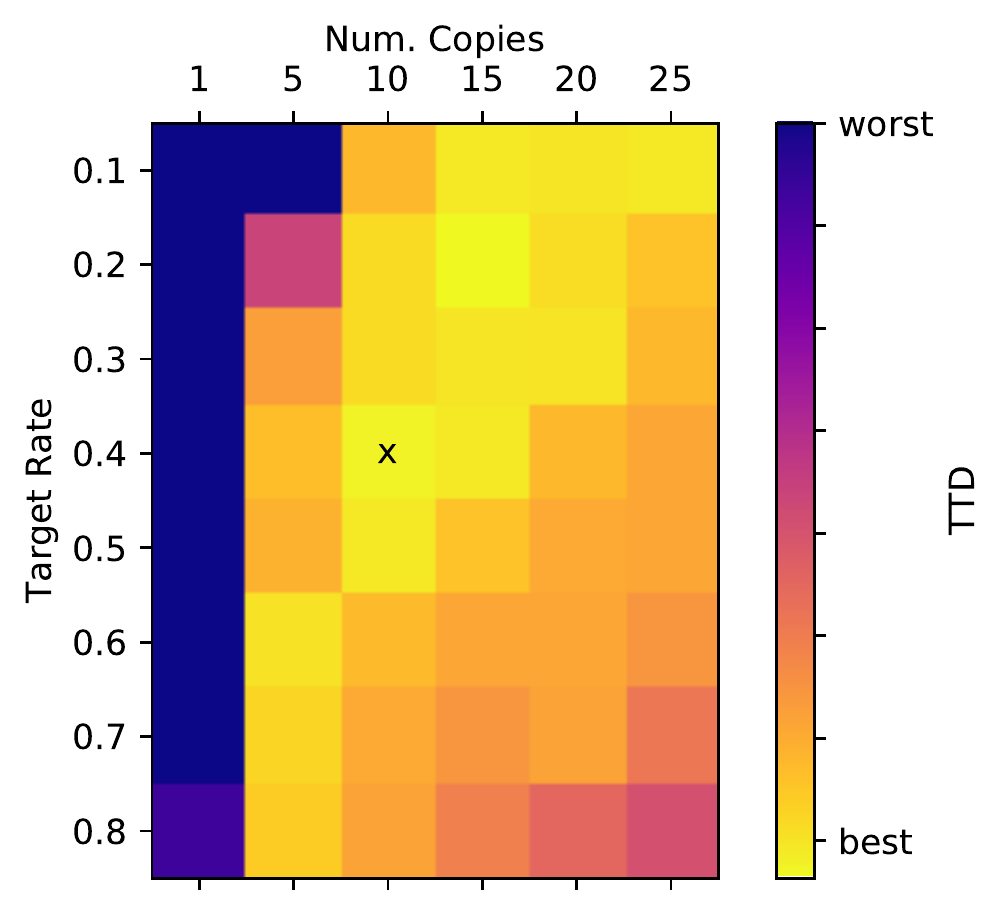}}
    \caption{Medians of TTD ($\alpha=0.01$, $R=0.25$) for a selection of problems on C12, for the hyper-parameters of PT (left panel) and PT+ICM (right panel). The ``x'' mark shows the optimal configuration of the hyper-parameters. Color coded in logarithmic scale with blue standing for largest TTD and yellow for smallest TTD. }
    \label{fig:pt_optimization}
\end{figure}

\begin{figure}
    \centering
    \includegraphics[width=.3\textwidth]{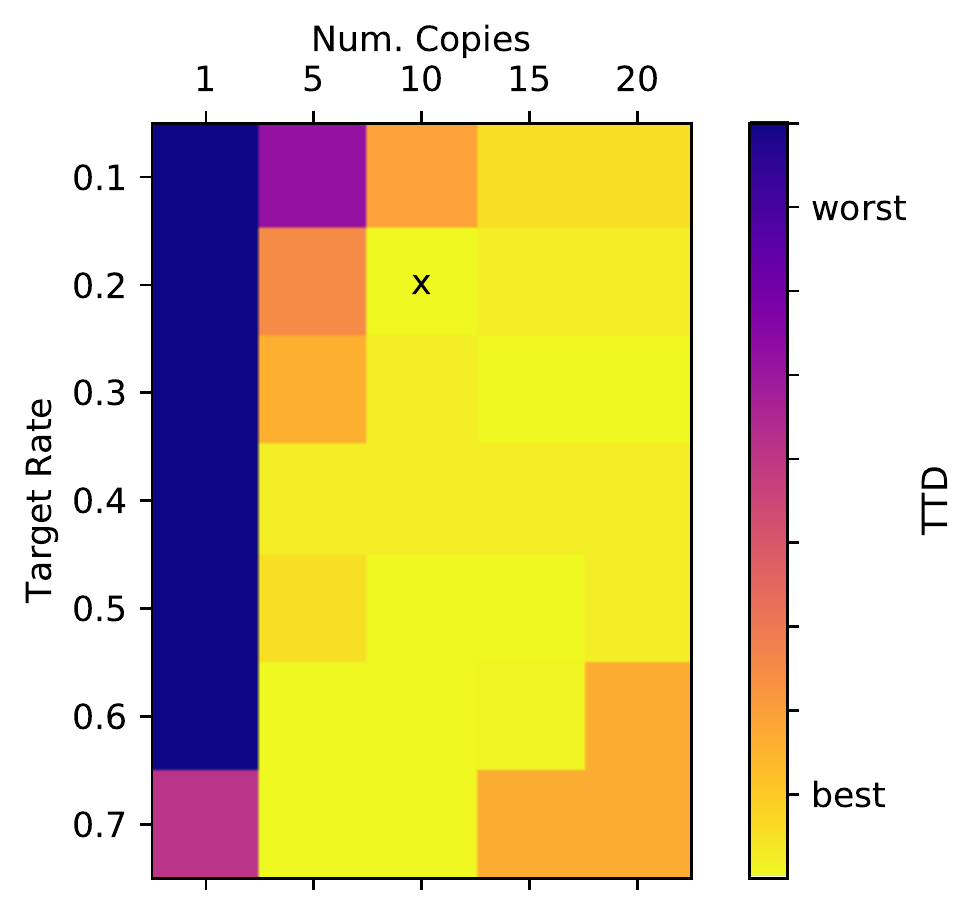}
    \caption{Medians of TTD ($\alpha=0.005$, $R=0.25$) for a selection of problems on C12, for the hyper-parameters of  PT+ICM. The ``x'' mark shows the optimal configuration of the hyper-parameters. Color coded in logarithmic scale with blue standing for largest TTD and yellow for smallest TTD. For this approximation ratio we did not consider the PT solver. With respect to the case in Fig.~\ref{fig:pt_optimization} the degeneracy between the two parameters is more pronounced.}
    \label{fig:pt_optimization_lower_alpha}
\end{figure}

For PT and PT+ICM we chose to optimize the commonly used target rate (the exchange rate between neighboring replicas) and the number of copies per temperature. For the former we adjusted the number and the spacing of the temperatures using the method of \cite{1999PhRvE..60.3606H}.  In our experiments we found beneficial to collect samples from multiple copies at the same temperature, see Fig.~\ref{fig:pt_config}. These can be seen as independent PT solvers running in parallel. However, there is a trade-off in the ``number of copies'' of PT running as while increasing them helps increasing the diversity of samples, the overhead of having multiple PTs considerably increases TTD. The results of PT optimizations are shown in Fig.~\ref{fig:pt_optimization}. We can see a degeneracy between target rate - proportional to the number of replicas - and number of copies of PT. This is expected as more copies of PT consist in an effective increase in number of replicas. Interestingly however, having only 1 copy of PT with larger target rate result in consistent larger TTD.

\subsection{Quantum Annealing}

\begin{figure}
    \centering
    \includegraphics[width=.35\textwidth]{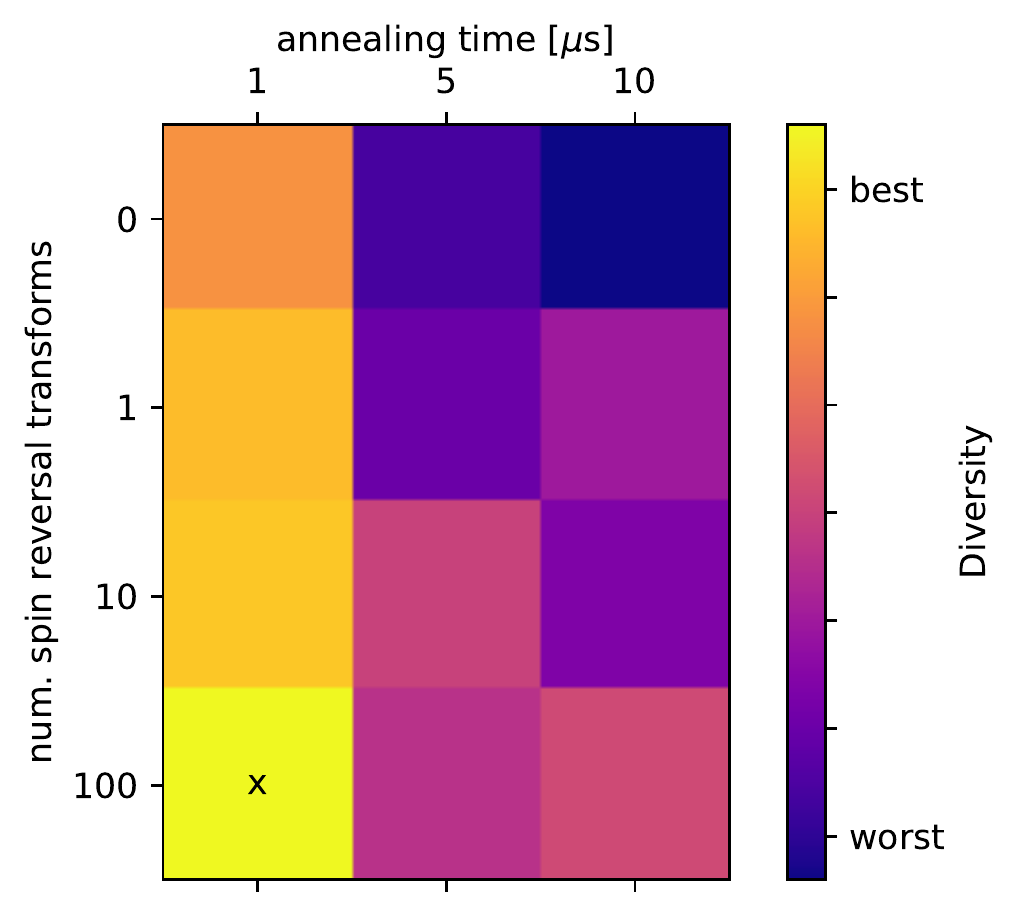}
    \caption{Optimization of quantum annealing. The colors show the total diversity after 50k samples: blue standing for low diversity and yellow for large diversity. The diversity is computed for an approximation ratio $\alpha=0.01$ and a radius factor $R=0.25$.}
    \label{fig:qa_optimization}
\end{figure}
For quantum annealing we optimized the annealing time and the number of spin reversal transformations. We chose a subset of the inputs and optimized overall diversity with approximation ratio $\alpha=0.01$ and radius factor $R=0.25$. Spin-reversal transformations are local transformations $\alpha = [\alpha_1, \dots, \alpha_{n}] \in \{\pm 1\}^{\times n}$ that flip the spin $i$ if $\alpha_i = -1$, accompanied by the Hamiltonian transformations
\begin{equation}
    h_i \to \alpha_i h_i, \quad J_{ij} \to \alpha_i \alpha_j J_{ij}.
\end{equation}
Although the Ising Hamiltonian \eqref{eqn:DWAVEHamiltoniainCS} does not change under these transformations, they can reduce systematic and analog errors on the QPU. We compute the total diversity of 50k samples obtained by the QPU with hyper-parameters on the grid $t_a \in [1 \, \mu{\rm s}, 5\, \mu{\rm s}, 10\, \mu{\rm s}]$ and number of spin reversal transformations (for each 10k samples batch) in [0, 1, 10, 100]. The results are shown in Fig.~\ref{fig:qa_optimization}.
\clearpage

\end{document}